\documentclass[preprint2]{aastex}

\usepackage{color}

\begin{document}

\title{Electron-ion scattering in dense multi-component plasmas:\\ application to the outer crust of an accreting neutron star}


\author{J. Daligault and S. Gupta}
\affil{Los Alamos National Laboratory, Los Alamos, New Mexico 87545}

\email{daligaul@lanl.gov,guptasanjib@lanl.gov}


\begin{abstract}
The thermal conductivity of a dense {\it multi-component} plasma is  critical to the modeling of accreting neutron stars.
To this end, we perform large-scale molecular dynamics simulations to calculate the static structure factor of the dense multi-component plasma in the neutron star crust from near the photosphere-ocean boundary to the vicinity of the neutron drip point.
The structure factors are used to validate a microscopic linear mixing rule that is valid for arbitrarily complex plasmas over a wide range of Coulomb couplings.
The microscopic mixing rule in turn implies and validates the linear mixing rule for the equation of state properties and also the linear mixing rule for the electrical and thermal conductivities of dense multi-component plasmas.
To make our result as useful as possible, for the specific cases of electrical and thermal conductivities, we provide a simple analytic fit that is valid for arbitrarily complex multi-component plasmas over a wide range of Coulomb couplings.
We compute the thermal conductivity for a representative compositional profile of the outer crust of an accreting neutron star in which hundreds of nuclear species can be present.
We utilize our results to re-examine the so-called impurity parameter formalism as used to characterize impure plasmas.
\end{abstract}


\keywords{Dense matter, nuclear reactions, nucleosynthesis, abundances, equation of state -- 
Stars; neutron --
X-rays; binaries, bursts}

\section{Introduction}

The thermal conductivity of a dense {\it multi-component} plasma is  critical to the modeling of the quiescent and explosive emission from the surface of accreting neutron stars. Much effort has been expended in accurately calculating the magnitude and location of crustal heating from nuclear reactions using modern nuclear models for energetics and reaction rates such as in \citet{HZ90,HZ03,Gupta2007,HZ08,Gupta2008}. However, crust thermal conductivity is of equal importance in determining the thermal structure of an accreting neutron star and the uncertainty in this quantity is compounded by the difficulties in understanding the phase diagram of a dense plasma with hundreds of nuclear species spread over several mass chains.

For illustration of the importance of thermal conductivity, we now describe two phenomena that have attracted much attention in the observational community: (i) superbursts and (ii) quasi-persistent transients.

(i) Long-term monitoring of Low-Mass X-ray Binaries (LMXBs) by instruments on board $BeppoSAX$ and $RXTE$ has yielded a wealth of phenomena on accreting neutron stars. 
For example, superbursts, lasting roughly a thousand times as long as Type-I X-ray bursts and recurring on timescales of a year, have energies $\sim 10^3$ times that of a typical X-ray burst. For the source 4U 1636-53 three superbursts were observed within 4.7 yr \citep{Wijnands2001, Kuulkers2004a}. These still mysterious explosions are thought to result from unstable $^{12}$C burning \citep{CB01} at column depths $\sim 10^{12}$ g cm$^{-2}$ (see \citet{Kuulkers2004b} for a review).
The crust thermal gradient dictates whether a temperature conducive to unstable $^{12}$C burning can exist at an ignition depth inferred from observations of superburst recurrence times.
A critical ingredient in predicting the thermal gradient that a neutron star crust can support is the thermal conductivity of the impure crust that is expected when Type-I X-ray bursts drive a rapid-proton-capture, or ``rp-process" on the neutron star surface.
Models that treat the crust thermal conductivity as uncertain also have to grapple with the large uncertainties in core neutrino luminosity from URCA processes. In this work we show that if the composition of the crust can be accurately computed, then so can crust thermal conductivity.

(ii) Another exciting application of accurate conductivities appears in modeling the cooling lightcurves of transiently accreting neutron stars. In the case of transients with accretion episodes of sufficient duration to heat the crust by deep crustal nuclear reactions (quasi-persistent transients), the thermal relaxation behavior gives us a new observational window into the physics of matter at very high density. Recently, \citet{Shternin2007}, \citet{Cackett2006} have presented detailed analyses of two such quasi-persistent transients, namely KS 1731-260 and MXB 1659-29. Since a critical determinant of the time evolution of surface flux is the thermal diffusion timescale as a function of depth in the neutron star, the conductivity plays an important role in modeling the cooling lightcurve of transients.

To reduce the uncertainties in modeling the thermal profiles of persistent and transient accretors, we have begun a program of accurately computing the microphysical quantities
that influence thermal structure. For the impure ashes from an X-ray burst, it is crucial to ascertain the composition evolution of these ashes with increasing depth in the neutron star crust. It is upon this composition profile that all other thermal structure determinants depend, most notably
\begin{itemize}
\item nuclear heating due to non-equilibrium electron captures, neutron reactions and pycnonuclear fusion, 
\item the thermal conductivity $\kappa$ and
\item the neutrino emissivity $\epsilon_{\nu}$ from Neutrino-Pair Bremsstrahlung (NPB).
\end{itemize}

The compositional profile of the outer crust (where there is negligible free neutron abundance) starting with a realistic {\it multi-component} plasma (MCP) of XRB ashes was calculated for the first time using global nuclear structure and reaction rate inputs to a large reaction network in \citep{Gupta2007}. By a ``composition profile" we refer to the changing composition from the top of the crust (X-ray burst ashes) all the way to neutron-drip. This spans four orders of magnitude in density (roughly $\sim10^{7.5}$ to $\sim10^{11.5}$ g/cc) allowing for a large range in Coulomb coupling. It is this compositional profile that we employ in the present work to investigate the resulting crustal microphysics since it has tens to hundreds of abundant species at each depth. Thus, we have an ideal site to explore the physics of strongly-coupled {\it multi-component} plasmas (MCP).
 
In this paper, we perform large-scale molecular dynamics (MD) simulations to calculate the structure factor of the dense MCP in the neutron star crust from near the photosphere-ocean boundary to the vicinity of the neutron drip point.
The structure factors are used to validate a microscopic linear mixing rule (MLMR) that is valid for arbitrarily complex MCPs over a wide range of Coulomb couplings.
The MLMR in turn implies and validates the mixing rule for the equation of state properties recently proposed for MCPs by \citet{PotekhinChabrierRogers}, and the mixing rule for the electrical and thermal conductivities of MCPs by \citet{Potekhin1999}.
These mixing rules provide a simple recipe to compute the physical properties of plasma mixtures, such as would be found in terrestrial experiments (e.g., inertial confinement fusion (ICF) plasmas.)
In general, the MLMR implies a linear mixing rule for any quantity that is an integral of the structure factor, such as the neutrino pair-Bremsstrahlung emissivity.
To make our result as useful as possible for the compact object modeling community, for the specific cases of electrical and thermal conductivities, we provide a simple analytic fit that is valid for arbitrarily complex MCPs over a wide range of Coulomb couplings.

In addtion, we comment on the standard characterization of crust impurity in terms of the impurity parameter $Q_{\rm imp}$, the mean-square deviation in species charge, and show that this is inadequate for the amorphous outer crust.
We compare the similarities and differences charaterizing electron-ion scattering in the crust of neutron stars and in terrestrial solid and liquid metals - we also comment on the controversy regarding the behavior of the electron-ion scattering at the solid-liquid phase boundary. 
We use the mixing rule to compute the thermal conductivity of the outer crust of an accreting neutron star as a function of depth. Since the X-ray burst ashes (initial crust composition) used to determine the crust composition profile are obtained from a model that has accretion rate $\sim 0.1 \dot{M}_{\rm Edd}$, where $\dot{M}_{\rm Edd} \approx 10^{18}$ g s$^{-1}$ is the Eddington accretion rate for a 10 km radius neutron star accreting a roughly solar hydrogen fraction, our results for crust conductivity are immediately applicable to superburst progenitors.

We restrict our study to the outer crust of the neutron star.
In the vicinity of neutron drip, \citet{Gupta2008} showed that a large number of reactions involving neutrons can remove compositional memory, and the complexity can be drastically reduced in the inner crust.
We plan to follow up on the nucleosynthesis results of \citet{Gupta2008} with a simple model of how an MCP evolves in the inner crust - this will serve as an analytic model of crust composition evolution on either side of neutron-drip without invoking large reaction networks.
Even in the scenario of unsuppressed photo-dissociation as investigated recently by \citet{Shternin_and_Yakovlev_2009}, the impurity of the outer crust is preserved from X-ray burst ashes on the surface to the rather high depth of $10^{11}$ g/cc. Between this depth and the neutron-drip point at $\sim 4\times 10^{11}$ g/cc, temperature-driven $(n,\gamma)-(\gamma,n)$ equilibria along isotopic chains will reduce the impurity to roughly $Q_{\rm{imp}}\sim 40$. Beyond neutron-drip (between $\sim 4 \times 10^{11}$ and $3 \times 10^{12}$ g/cc) the density-dependent electron-capture-delayed neutron emissions, (EC,xn), as described in \citet{Gupta2008}, will be the primary drivers of abundance re-arrangement and crust impurity will be reduced very quickly with increasing depth. Between $\sim 3 \times 10^{12}$ and $\sim 10^{13}$ g/cc pycnonuclear reactions will change the composition using the few remaining nuclides to determine the compositional trajectory. Finally, at densities greater than $10^{13}$ g/cc proton-tunneling reactions will play a role, and may be instrumental in driving the composition to a single species \citep{Jones2005} without diverse metastable state populations spread out over different proton shell closures.
The abrupt change in composition and resulting change in the phase diagram at neutron drip will be addressed in a later paper.
Therefore this paper is restricted to the highly impure liquid metallic and amorphous phases of the neutron star crust which exist at depths shallower than neutron drip. 

\section{Electronic properties of ionic mixtures} \label{section2}

The section is organized as follows:
\begin{itemize}
\item First, we briefly discuss the principal charateristics of the MCPs found in the crust of accreting neutron star;
\item we then recall the link between the electronic transport properties and NPB emissivity with the charge-charge structure factor;
\item we discuss the standard treatments of mixtures, namely the impurity parameter formalism and the linear mixing rule;
\item finally, we compare electron scattering in terrestrial metals and neutron star crusts, and comment on the model of electron-ion scattering at the solid-liquid phase boundary as proposed by \citet{Baiko1998}.
\end{itemize}

All the formulas below are expressed in natural (SI) units (in particular $e^{2}=q^{2}/4\pi\epsilon_{0}$ where $q$ is the electronic charge.)
Also, the words ``nucleus (nuclei)" and ``ion (ions)" are used interchangeably.

\subsection{Characterizing the ionic mixtures}

We consider an unmagnetized, neutral {\it multi-component} plasma (MCP) in a volume $V$ at temperature $T$ consisting of a mixture of $N_{\rm sp}$ atomic species of charge $Z_{j}$ and mass number $A_{j}$, $j=1,...,N_{\rm sp}$.
The number and mass densities of species $j$ are $n_{j}$ and $\rho_{j}$, respectively.
The total number density and mass density are $n=\sum_{j=1}^{N_{\rm sp}}{n_{j}}$ and $\rho=\sum_{j=1}^{N_{\rm sp}}{\rho_{j}}$, respectively.
The total electron density is $n_{\rm e}=\sum_{j}^{N_{\rm sp}}{Z_{j}n_{j}}=\langle Z\rangle n$, where $\langle Z\rangle=\sum_{j}^{N_{\rm sp}}{x_{j}Z_{j}}$ is the mean ionic charge.
We use 
\begin{eqnarray}
x_{j}=n_{j}/n
\label{numberfraction}
\end{eqnarray}
to denote the \emph{number fraction} of the $j$th species. Note that the number fraction is related to the species ``abundances" $Y_j$ as traditionally used in the astrophysics literature, therefore number fraction $x_j$ differ from the usual ``mass fraction" $X_j$.
Thus
\begin{eqnarray}
n_j= \rho N_A  Y_j = \rho N_{\rm A} \left(  \frac{X_j}{A_j}  \right)
\label{abundance}
\end{eqnarray}
where $A_j$ is the mass number of species $j$, $N_{\rm A}$ is Avogadro's number and
\begin{eqnarray}
x_j = \frac{X_j/A_j}{\sum_k (X_k/A_k)}
\label{numberfractionmassfraction}
\end{eqnarray}
where the index $k$ denotes summation over all nuclear species in the MCP, including free nucleons.
The average inter-particle distance between nuclei is denoted by $a=(3/4\pi n)^{1/3}$, whereas the mean distance between electrons is $a_{\rm e}=(3/4\pi\,n_{\rm e})^{1/3}$.
We will occasionally use the compact notations $\vec{Z}=(Z_{1},\dots,Z_{\rm N_{\rm sp}})$ and $\vec{x}=(x_{1},\dots,x_{\rm N_{\rm sp}})$ to denote the composition.
When $N_{\rm sp}=1$ the system is referred to as an one-component plasma (OCP) and we drop the subscript referring to the particles' species.

We define several dimensionless parameters to characterize the system, namely:
\begin{enumerate}
\item the relativistic parameter $x_{\rm r}=p_{\rm F}/m_{\rm e}c$; 
\item the degeneracy parameter $\Theta=T/T_{\rm F}$;
\item the electron coupling parameter $\Gamma_{\rm e}=e^{2}/a_{\rm e}k_{\rm B}T$,
\end{enumerate}
where $p_{\rm F}=\hbar k_{\rm F}$ is the Fermi momentum, $k_{\rm F}=(3\pi^{2}n_{\rm e})^{1/3}$ is the Fermi wavevector and $T_{\rm F}$ is the Fermi temperature of electrons.

For the typical physical conditions of interest in this paper, the density is in the range $10^{6}\,{\rm g.cm^{-3}}\leq\rho\leq 10^{12}\, {\rm g.cm^{-3}}$, the temperature is $T\sim 0.5$ GK, and the number of species $N_{\rm sp}$ varies from a few to several hundreds.
As illustrated in Fig.\ref{figure1}, at these densities and temperatures, $x_{\rm r}>1$, $\Theta<<1$ and $a/a_{\rm B}<<1$ where $a_{\rm B}$ is the Bohr radius.
As a consequence, atoms are completely pressure-ionized and matter is a dense, {\it multi-component} plasma composed of bare nuclei neutralized by a relativistic and strongly degenerate gas of delocalized electrons.
This paper deals with the scattering of those electrons off the {\it multi-component} ionic mixture.

The ions are in general non-relativistic - we shall also assume hereafter that the ions behave as classical entities, thereby neglecting the quantum nature of the ionic density fluctuations that may occur in the regions where $T>T_{\rm p}=\hbar\omega_{\rm p}/k_{\rm B}$, where $\omega_{\rm p}$ is the (kinetic) ion plasma frequency.

The nuclei are ``strongly coupled" in the sense that their properties are primarily governed by their mutual interactions rather than by purely thermal effects as in an ideal gas.

In the OCP, electron-ion scattering is fairly well understood. The degree of coupling of an OCP is usually characterized by the so-called Coulomb coupling parameter $\Gamma_{\rm OCP}=Z^{2}e^{2}/ak_{\rm B}T$, the ratio of the mean interaction energy $Z^{2}e^{2}/a$ to the thermal energy $k_{\rm B}T$ of the particles.
Neglecting electronic screening, the OCP is known to crystallize into a bcc lattice at $\Gamma_{\rm OCP}=\Gamma_{\rm c}\approx 175$.
Below $\Gamma_{\rm c}$ the OCP is in a fluid state whose properties vary from gas-like at small $\Gamma_{\rm OCP}<0.1 $ to liquid-like when $50<\Gamma_{\rm OCP}<\Gamma_{\rm c}$ \citep{Daligault2006}; in the intermediate regime, $0.1 < \Gamma_{\rm OCP}<50$, both potential and kinetic effects play a comparable role in a non-trivial way.
As a consequence, models of neutron star crusts based on the OCP usually treat the electron-ion scattering as follows:
\begin{itemize}
\item when $\Gamma>\Gamma_{\rm c}$, electrons travel in a bcc crystal and electron-ion scattering is interpreted as the scattering of Bloch electrons off the thermal vibrations (phonons) of the ions with respect to the bcc lattice sites.
\item when $\Gamma<\Gamma_{\rm c}$, electrons scatter off the disordered ionic background characteristic of the fluid state.
\end{itemize}
Recently \citet{Baiko1998} called into question this canonical picture because it yields discontinuities in the electronic transport properties at melting.
In section \ref{discontinuity}, we comment on this controversy regarding the behavior of the electron-ion scattering at the solid-liquid phase boundary of an OCP.

The state of affairs for {\it multi-component} mixtures is much less known, and especially for mixtures composed of $N_{\rm sp}\geq 3$ species as encountered in the crust of accreting neutron stars.
A common approach used to infer the properties of a mixture is to map the latter onto an effective OCP of coupling $\Gamma_{\rm eff}$, thereby disregarding the (poorly-understood) complexities of the phase diagram and dynamic properties of a true mixture.
A suitable definition of the effective coupling is
\begin{eqnarray}
\Gamma_{\rm eff}&=&\sum_{j=1}^{N_{\rm sp}}{x_{j}\Gamma_{j}}\nonumber\\
&=&{\langle Z^{5/3}\rangle}\Gamma_{\rm e} \label{Gamma}
\end{eqnarray}
where 
\begin{eqnarray}
\Gamma_{j}=(Z_{j}^{2}e^{2})/(a_{j}k_{\rm B}T) \label{Gammaj}
\end{eqnarray}
is the coupling parameter for species $j$, $a_{j}=(3/4\pi n_{j})^{1/3}$ being the average inter-particle distance of particles of species $j$,  and $\langle Z^{5/3}\rangle=\sum_{j}^{N_{\rm sp}}{x_{j}Z_{j}^{5/3}}$.
This choice for $\Gamma_{\rm eff}$ is motivated by the inequality (written here for pure Coulomb interactions) \citep{Rosenfeld1993}
\begin{eqnarray}
-\frac{9}{10}\Gamma_{\rm eff} \leq \frac{U_{\rm ex}/nV}{k_{\rm B}T}\/,
\end{eqnarray}
which provides a very good lower bound of the ratio of mean excess (interaction) energy per particle $U_{\rm ex}/nV$ to the thermal energy $k_{\rm B}T$, and therefore is a lower bound on the effective Coulomb coupling in the MCP.

In the astrophysical literature related to dense plasmas, often a mixture is assumed to crystallize at $\Gamma_{\rm eff}=175$ as in the OCP.
Several questions can be raised with regard to the onset of periodic ordering in a true MCP that have not been satisfactorily addressed to date.
For instance:
\begin{itemize}
\item[(a)] Does the system crystallize in a regular, periodic lattice as has often been assumed in the literature? 
\item[(b)] Do the most abundant species make a lattice and the less abundant species contribute to defects in the structure?
\item[(c)] Is it amorphous with a liquid-like frozen-in structure?
\end{itemize}
An affirmative answer to questions (a) and (b) above is assumed rather commonly in the literature and consequently, beyond $\Gamma_{\rm eff} > \Gamma_{\rm c}$, the total electron-ion scattering cross-section is split into two contributions: (1) the electron-phonon contribution due to the lattice structure, and (2) the electron-impurity contribution.
We believe that the phase diagram of mixtures such as those existing in the crust of accreting neutron stars is still too poorly known to make any definite statements such as (a) and (b) above.
The local composition may also be changed due to phase-separation (demixing) and sedimentation effects, which are still not well understood and are being investigated \citep{Horowitz2008}.
Given the complexity of the mixtures, it is, however, not unreasonable to think that, even when $\Gamma_{\rm eff}>\Gamma_{\rm c}$, the system will most likely be \emph{amorphous}, or at the very least be comprised of many defects, and that the existence of Bloch states is questionable.
In the following, we shall assume that the composition at any depth in the outer crust of the accreting neutron star is known; in practice, we shall use the outer crust composition from \citet{Gupta2007}.
In all cases that we considered, when starting from an amorphous structure, the molecular dynamics simulations performed did not show any evidence for demixing or crystallization over the timescale of the simulations.

\subsection{Electron-ion scattering in MCPs}

 We shall focus on electron-ion scattering and ignore the effect of electron-electron ($ee$) scattering.
It was shown in \citep{Potekhin1999} that $ee$ scattering contributes to conductivity only for 
very light nuclei with $Z \lesssim 6$ and under non-degenerate conditions. For a pure $^{56}$Fe
composition in a degenerate plasma at densities $\approx 10^{8.5}-10^{11.5}$ g/cc \citep{Potekhin1999} showed 
that $ee$ collisions could be neglected.

Many of the basic formulas in the following sections can be found, for instance, in \citet{FlowersItoh1976}.

\subsubsection{Thermal and Electrical Conductivities}

Electron-ion scattering impedes the transport of electronic momentum and energy.
The resulting electrical conductivity $\sigma$ and the thermal conductivity $\kappa$ can generally be expressed in terms of the collision frequencies $\nu_{\sigma}$ and $\nu_{\kappa}$ :
\begin{eqnarray}
\sigma&=&\frac{n_{\rm e}e^{2}}{m_{\rm e}^{*}}\tau_{\sigma}\\
\kappa&=&\frac{1}{3}v_{\rm F}^{2}C_{\rm V}\tau_{\kappa}=\frac{\pi^{2}k_{\rm B}^{2}T\/n_{\rm e}}{3m_{\rm e}^{*}}\tau_{\kappa}\,,
\end{eqnarray}
where $m_{\rm e}^{*}=m_{\rm e}\sqrt{1+x_{\rm r}^{2}}$ is the effective electron mass, $v_{\rm F}=p_{\rm F}/m_{\rm e}^{*}$ is the Fermi velocity, and $C_{\rm V}$ is the heat capacity of the free relativistic electron gas.

The quantities $\tau_{\sigma}$ and $\tau_{\kappa}$ describe the typical timescale over which electron-ion collisions degrade the electric and thermal currents.
Their inverses, the frequencies $\nu_{\sigma,\kappa}=1/\tau_{\sigma.\kappa}$, can conveniently be written as the product of a ``fundamental'' frequency $\nu_{0}$:
\begin{eqnarray}
\nu_{0}=\frac{4\pi \langle Z^{2}\rangle e^{4}n_{i}}{p_{\rm F}^{2}v_{\rm F}}=\frac{4\alpha^{2} E_{\rm F}}{3\pi\hbar}\frac{\langle Z^{2}\rangle}{\langle Z\rangle}\/,
\label{fundfreq}
\end{eqnarray}
multiplied by a dimensionless quantity $\ln\Lambda_{\sigma,\kappa}$ known as the Coulomb logarithm:
\begin{eqnarray}
\nu_{\sigma,\kappa}=\nu_{0}\ln\Lambda_{\sigma,\kappa}\/. \label{nusigmakappa}
\end{eqnarray}
In Eq.(\ref{fundfreq}), $\langle Z^{2}\rangle$ results form the appropriate normalization $S_{\rm zz}(k\to\infty)\to 1$ of the structure factor discussed below in Eq.(\ref{Szz}), $\alpha=e^{2}/\hbar c$ is the fine-structure constant and $E_{\rm F}=m_{e}^{*}c^{2}$ is the electron Fermi energy.
The fundamental frequency $\nu_{0}$ can be seen as a frequency of Coulomb collisions associated with the transfer of momentum between electrons and ions; it can be written as
\begin{eqnarray}
\nu_{0}=n\left[4\pi \left(\frac{\langle Z^{2}\rangle^{\frac{1}{2}}e^{2}}{m_{e}^{*}v_{\rm F}^{2}}\right)^{2}\right]v_{\rm F}\/,
\end{eqnarray}
in which the bracketed term can be interpreted as the cross-section for momentum transfer in a binary collision between an electron of velocity $v_{\rm F}$ and charge $-e$ and a massive ion of charge $\langle Z^{2}\rangle^{\frac{1}{2}}e$. 
The dimensionless Coulomb logarithm contains all the fine details of electron-ion collisions in a plasma, such as charge screening, collective modes and quantum diffraction effects.
In general, $\ln\Lambda_{\sigma}\neq \ln\Lambda_{\kappa}$ since the two transport coefficients measure different mechanisms, namely momentum and energy transport.
However, when the dominant source of electron-ion scattering is elastic, as in the disordered fluid phase or the high-temperature crystalline phase, the two Coulomb logarithms are equal and the Wiedemann-Franz relation 
\begin{eqnarray}
\label{WF}
\kappa/\sigma=(\pi^{2}k_{B}^{2}/3e^{2})T
\end{eqnarray}
is satisfied.

A calculation of the Coulomb logarithm that self-consistently treats all the many-body effects occuring in a multi-component plasma is in general rather complicated.
However, because of the high electron density in neutron stars, the typical electron kinetic energy is very large compared to the electron-ion interaction energy and the average electron-ion correlations are weak.
Accordingly, a good approximation can be made by considering the ions and electrons in the plasma as two weakly interacting subsystems, namely an MCP neutralized by a homogeneous negatively-charged background and an homogeneous interacting electron gas (i.e. relativistic quantum electronic jellium.)
Electrons scatter off the electrostatic potential created by the ionic charge distribution,
\begin{eqnarray} \label{rhoz}
n_{z}({\bf r},t)=\sum_{j=1}^{N_{\rm sp}}{Z_{j}\sum_{l=1}^{N_{j}}{\delta\left({\bf r}-{\bf R}_{l}(t)\right)}}\/,
\end{eqnarray}
where ${\bf R}_{l}(t)$ is the position of the $l$-th nuclei of the $j$-th species at time $t$, and $N_{j}=n_{j}V$ is the number of particles of species $j$.
In the (first) Born approximation\footnote{Extensions to higher-order approximations are difficult.
An approximate treatment proposed by \citet{Yakovlev1987} prescribes usage of the exact binary cross-section instead of the Born approximation in the integrand of Eq.(\ref{CoulombLogarithm}).
This introduces an additional factor in Eq.(\ref{CoulombLogarithm}), namely $R(k)=\sigma(k)/\sigma_{B}(k)$ where $\sigma$ is the exact differential cross-section for a momentum transfer $k$ and $\sigma_{B}$ is its Born approximation.
A similar procedure can be applied verbatim to our calculations since the MCP structure factor is not affected.
Recently, a similar averaging procedure was used by \citet{Itoh2008}, however the title of that paper is somewhat misleading since it is not actually the second-Born approximation that is performed therein. Such a calculation would involve the higher-order correlation functions of the ionic density Eq.(\ref{rhoz}). However, these higher-order correlations are noticeably absent in the equations of \citet{Itoh2008} for the transport coefficients.}
 of the electron-ion interaction and using the fact that the electrons are much heavier than the ions, we obtain \citep{FlowersItoh1976}:
\begin{eqnarray} 
\lefteqn{\ln\Lambda_{\sigma,\kappa}=}&&\label{CoulombLogarithm}\\
&&\int_{0}^{2k_{\rm F}}{dk\/k^{3}\Big|\frac{v(k)}{\epsilon_{e}(k)}\Big|^{2}\left[1-\frac{x_{\rm r}^{2}}{1+x_{\rm r}^{2}}\left(\frac{k}{2k_{\rm F}}\right)^{2}\right]{\cal{S}_{\sigma,\kappa}}(k)}\/,\nonumber
\end{eqnarray}
with $v(k)=1/k^{2}$ and $\epsilon_{e}(k)$ is the static dielectric function of the electron jellium.

Equation (\ref{CoulombLogarithm}) can be interpreted as follows.
The Coulomb logarithm $\ln\Lambda_{\sigma,\kappa}$ sums all the contributions of collisions characterized by momentum transfer $k$.
The range of integration is limited to $2k_{\rm F}$ because the electrons are degenerate and only those electrons near the Fermi surface take part in exchanges of momentum and energy.
The electrons behave as if they were scattered independently of each other by the electronically screened electron-ion potential.
Because the electrons are much lighter than the nuclei, the energy exchanges $\hbar\omega$ in inelastic collisions occur in a range over which $\epsilon_{e}(k,\omega)\approx\epsilon_{e}(k,0)=\epsilon_{e}(k)$.
For relativistic, degenerate electrons, an analytical expression for $\epsilon_{e}$ was obtained by \citet{Jancovici1962} within the Random-Phase Approximation (see Appendix \ref{appendix2}).

The bracketed term, which describes the kinematic suppression of backward scattering of relativistic electrons, $\left[ 1-\frac{x_{\rm r}^{2}}{1+x_{\rm r}^{2}}\left(\frac{k}{2k_{\rm F}}\right)^{2} \right]$, varies between unity for $x_{\rm r}\to 0$ and $\left[ 1-(k/2k_{\rm F})^{2} \right]$ for $x_{\rm r}\to\infty$.

The many-body physics of the ionic subsystem is encapsulated in the function ${\cal{S}_{\sigma,\kappa}}(k)$.
${\cal{S}_{\sigma,\kappa}}(k)$ is a functional of the spectrum of charged fluctuations $\delta n_{z}({\bf k},t)=n_{z}({\bf k},t)-\langle n_{z}({\bf k})\rangle_{\rm eq}$ in the ionic system
\begin{eqnarray}
\lefteqn{S_{\rm zz}(k,\omega)=}&& \label{Szz}\\
&&\frac{1}{nV\langle Z^{2}\rangle}\int{\frac{d\omega}{2\pi}\left\langle\delta n_{z}({\bf k},t)\,\delta n_{z}(-{\bf k},0)\right\rangle_{\rm eq} e^{i\omega\/t}}\nonumber
\end{eqnarray}
where $\langle\,\rangle_{\rm eq}$ denotes the average of a canonical ensemble at temperature $T$.
In other words, ${\cal{S}_{\sigma,\kappa}}(k)$ self-consistently takes into account all of the separate electron-ion relaxation mechanisms at the level of the Born approximation; it is really an umbrella term for electron-ion, electron-impurity, electron-phonon scattering and includes multi-phonon processes.
We have, from \citet{FlowersItoh1976},
\begin{eqnarray*} 
\!\!\!{\cal{S}}_{\sigma}(k)&=&\int_{-\infty}^{\infty}{S_{\rm zz}(k,\omega)z\/n(z)} \quad z=\hbar\omega/k_{B}T\\
\!\!\!{\cal{S}}_{\kappa}(k)&=&\int_{-\infty}^{\infty}{S_{\rm zz}(k,\omega)z\/n(z)\left[1+\left(\frac{3k_{\rm F}^{2}}{k^{2}}-\frac{1}{2}\right)\right]z^{2}}
\end{eqnarray*}
where $n(z)=1/1-e^{-z}$.

In the special case of a periodic, crystalline structure, the term $\langle n_{z}({\bf k})\rangle_{\rm eq}$ peaks at values of the wavevector ${\bf k}$ commensurate with the periodicity of the lattice and substracts out the elastic Bragg peaks: electron-ion scattering involves Bloch electrons and Eq.(\ref{CoulombLogarithm}) is obtained by approximating their behavior by means of free particle wave functions.

When the scattering is predominantly elastic in nature, as is expected in the fluid and solid phases at $T>T_{p}$ or in a disordered (non-periodic) lattice, we have
\begin{eqnarray} 
{\cal{S}}_{\sigma,\kappa}(k)&=&\int_{-\infty}^{\infty}{S_{\rm zz}(k,\omega)}\\
&=&\frac{1}{nV\langle Z^{2}\rangle}\left\langle\delta n_{z}({\bf k},0)\,\delta n_{z}(-{\bf k},0)\right\rangle_{\rm eq}\\
&\equiv&S_{\rm zz}(k) \label{Sqq}
\end{eqnarray}
where $S_{\rm zz}(k)$ is the (charge-charge) structure factor. This leads to
\begin{eqnarray} 
\lefteqn{\ln\Lambda_{\sigma,\kappa}=}&&\\
&&\int_{0}^{2k_{\rm F}}{dk\/k^{3}\Big|\frac{v(k)}{\epsilon(k)}\Big|^{2}\left[1-\frac{x_{\rm r}^{2}}{1+x_{\rm r}^{2}}\left(\frac{k}{2k_{\rm F}}\right)^{2}\right]S_{\rm zz}(k)}\/.\nonumber
\label{coul_log_szz}
\end{eqnarray}
In this limit, $\sigma$ and $\kappa$ are insensitive to the details of the spectrum of ionic charge fluctuations.
These details (ion collective modes and single-particle effects) are present implicitly but in an integrated form via the sum-rule (\ref{Sqq}).

Until now, we have assumed that the ions are point particles.
In this ``point-charge" approximation, the only species-dependent term in the integrand of Eq.(\ref{CoulombLogarithm}) is the charge-charge structure factor $S_{\rm zz}(k)$. 
In order to allow for finite nuclear size, one can replace the density (\ref{rhoz}) with
\begin{eqnarray}
n_{z}({\bf k},t)=\sum_{j=1}^{N_{\rm sp}}{Z_{j}F_{j}(k)\sum_{l=1}^{N_{j}}{e^{-i{\bf k}\cdot{\bf R}_{l}(t)}}}\/.
\end{eqnarray}
where $F_{j}(k)$ is the nuclear form factor, which reflects the charge distribution within a nucleus of species $j$.
For instance, assuming uniform proton density in the nuclei,
\begin{eqnarray}
F_{j}(k)=\frac{3}{(k\/R_{j})^3}\left(\sin(k\/R_{j})-k\/R_{j}\cos(k\/R_{j})\right) \label{formfactor}
\end{eqnarray}
where $R_{j}$ the charge radius of the nuclear species $j$ which has mass number $A_{j}$; we  use $R_{j}=1.15A_{j}^{1/3}$ fm as in \citet{KaminkerNPB1999}.
Direct MD calculation of the structure factor $S_{\rm zz}(k)$ with and without the nuclear form factor (\ref{formfactor}) shows that the finite nuclear size corrections are completely negligible in the outer crust of interest in this paper.
(Note that for instance $(F_{j}(k))^2=0.9766$ at the peak of $S_{\rm zz}(k)$, i.e. at $(k/2k_F) \approx 0.33$, for a typical heavy nucleus having mass number $A_j\approx64-100$ or $A_{j}^{1/3} \approx 4 - 4.6$).
Accordingly we neglect finite nuclear size effects in the remainder of this paper.

\subsubsection{Neutrino-pair Bremsstrahlung (NPB) emissivity}
In NPB, the electrons are accelerated by the Coulomb field of the crust as in photon Bremsstrahlung but with the emission of a neutrino-antineutrino pair instead of a photon.
In the first Born approximation, the neutrino emissivity is
\begin{eqnarray}
\epsilon_{\nu}=\frac{8\pi G_{\rm F}\langle Z^{2}\rangle e^{4}C_{+}^{2}}{567\hbar^{9}c^{8}}(k_{\rm B}T)^{6}\ln\Lambda_{\nu}
\end{eqnarray}
with the Coulomb logarithm for NPB,
\begin{eqnarray}
\ln\Lambda_{\epsilon}&=&\frac{\hbar c}{k_{\rm B}T}\int_{0}^{2k_{\rm F}}{dk_{\rm t} k_{\rm t}^{3}\int_{0}^{\infty}{dk_{\rm r}\Big|\frac{v(k)}{\epsilon(k)}\Big|^{2}S_{\rm zz}(k)R_{\rm T}(k_{\rm t},k_{\rm r})}} \nonumber\\
&\approx&\int_{0}^{2k_{\rm F}}{dk\/k^{3}\Big|\frac{v(k)}{\epsilon(k)}\Big|^{2}a(k)S_{\rm zz}(k)}\label{CoulombLogarithmNPB}
\end{eqnarray}
where ${\bf k}={\bf k}_{\rm t}+{\bf k}_{\rm r}$, $G_{\rm F}$ is the Fermi weak coupling constant, $C_{+}^{2}\approx 1.675$ from the vector and axial-vector constants of the weak interaction and $a(k)= \left[ 1-\frac{2k^{2}}{4k_{\rm F}^{2}-k^{2}}\ln\left(\frac{k}{2k_{\rm F}}\right) \right]$.
The function $R_{\rm T}$ describes the thermal broadening of the electron Fermi surface and from Eq.(20) of \citet{Haensel1996} this can be approximated as unity for $T \lesssim 0.6 $ GK typical of models of the neutron star crust at accretion rates characteristic of superburst progenitors \citep{KaminkerNPB1999}.

Thus, in both Eq.(\ref{CoulombLogarithm}) and Eq.(\ref{CoulombLogarithmNPB}) the only dependence on the nuclear species present (i.e., on the composition) enters via $S_{\rm zz}(k)$. Therefore, to tabulate fits to both $\kappa$ and $\epsilon_{\nu}$ for every nuclear species prior to application of a mixing rule is extremely cumbersome (note that the papers \citep{Itoh2008,ItohNPB1996} tabulate fit coefficients for only 11 species for conductivities and NPB emissivities respectively) and also unnecessary from a computational standpoint. The fits to OCP structure factors in \citet{Young} are sufficient to cover every one of the $N_{\rm sp} > 100$ species encountered in the neutron star crust. They can also be used to compute both $\kappa$ and $\epsilon_{\nu}$ for an MCP of arbitrary complexity using the microscopic mixing rule (MLMR)  described in section \ref{mixingrule} which involves only the OCP structure factors and species abundances. For higher values of $\Gamma > 225$ in the neutron star crust we present a simple model for the behavior of the Coulomb logarithm that reduces the evaluation of thermal and electrical conductivity to an analytic formula incorporating crust impurity in a very simple way. We do not present calculations of the NPB neutrino emissivity in this paper, rather we point it out to demonstrate that all such ``integral" determinants (i.e. summations over momenta) of crust thermal profile are related to the crust composition via the quantity $S_{\rm zz}(k)$. 

\subsubsection{The standard treatment of mixtures in the crust}

For simplicity, we drop the $\sigma$, $\kappa$ and $\nu$ subscripts in what follows.

The most common approach of electron-ion scattering in the neutron star crust relies on the belief that the MCP is fluid in the upper part of the crust where $\Gamma_{\rm eff}<175$ and consist of a solid, i.e. a crystalline lattice with impurities, in the lower part where $\Gamma_{\rm eff}>175$.
Accordingly, in the solid phase, the collision frequency $\nu=\nu_{\rm e-ph}+\nu_{\rm e-imp}$ is split into the sum of the electron-phonon and the electron-impurity scattering contributions; in the liquid state $\nu=\nu_{\rm ei}$ where $\nu_{\rm ei}$ is the electron-ion scattering inverse relaxation time discussed in the next section.

The splitting $\nu=\nu_{\rm e-ph}+\nu_{\rm e-imp}$ was initially introduced in \citet{FlowersItoh1976} to deal with periodic crystals with a small fraction of the lattice sites occupied by impurities.
The idea is illustrated in Fig.(\ref{figure2}).
The case of binary mixtures ($N_{\rm sp}=2$) was discussed in detail in \citet{ItohKohyama1993} for a small concentration of impurity ions (say species ``2"):  the contribution of the impurity to electron-ion scattering is equal to its concentration times the binary electron-ion collision frequency between an electron and a ``residual'' ion of charge $(Z_{2}-Z_{1})$.
This idea was extended to deal with more complex MCPs containing $N_{\rm sp}\geq 3$ species, in which case the system is treated as a perfect crystal of charge $\langle Z\rangle$ plus the contribution of {\it uncorrelated} impurities of charge $(Z_{\rm i}-\langle Z\rangle)$.

Within this approximation and defining the impurity parameter as
\begin{eqnarray}
Q_{\rm imp}=\sum_{j=1}^{N_{\rm sp}}{x_{i}(Z_{i}-\langle Z\rangle)^{2}}\/, \label{Qimp}
\end{eqnarray}
the relaxation frequency can be written as
\begin{eqnarray}
\label{nuimp}
\nu_{\rm imp}=\nu_{0,\rm imp}\ln\Lambda_{\rm imp}
\end{eqnarray}
where
\begin{eqnarray}
\nu_{0,\rm imp} = \frac{4\alpha^{2}\epsilon_{\rm F}}{3\pi\hbar}\frac{Q_{\rm imp}}{\langle Z\rangle}
\end{eqnarray}
and
\begin{eqnarray}
\ln\Lambda_{\rm imp}=\int_{0}^{2k_{\rm F}}{dk\/k^{3}\Big|\frac{v(k)}{\epsilon(k)}\Big|^{2}\left[1-\frac{x_{\rm r}^{2}}{1+x_{\rm r}^{2}}\left(\frac{k}{2k_{\rm F}}\right)^{2}\right]}\/. \label{lambdaimp}
\end{eqnarray}
$\ln\Lambda_{\rm imp}$ corresponds to Eq.(\ref{coul_log_szz}) with $S_{\rm zz}$ set to unity to correspond to independent (uncorrelated) electron-impurity collisions.

While the impurity parameter formalism was developed to treat small impurity concentration in an otherwise almost perfect crystal lattice (see Fig.(\ref{figure2}), it has also used to characterize electron-ion scattering in complex fluid and amorphous MCPs when $Q_{\rm imp}$ is large, say $Q_{\rm imp}\sim 100$.
Such is usually the case when electron-impurity scattering is used to set a lower bound on conductivity for the entire crust.
However, since $\Lambda_{\rm imp}$ is of the order of magnitude as $\Lambda$, we have
\begin{eqnarray}
\nu/\nu_{\rm imp}&=&\frac{\langle Z^{2}\rangle}{Q_{\rm imp}}\frac{\ln\Lambda}{\ln\Lambda_{\rm imp}}\\
&\approx& \frac{\langle Z^{2}\rangle}{Q_{\rm imp}}\sim 10
\end{eqnarray}
over the whole outer crust as shown in Fig.\ref{figure1}, the impurity parameter can lead to conductivities much higher than the those that would be obtained using the actual structure factors employed in Eq.(\ref{coul_log_szz}).
We thus anticipate that $Q_{\rm imp}$ is inappropriate for the outer crust (impure liquid metal approaching heterogeneous amorphous solid at high density), and instead we shall justify the usage of $\langle Z^{2}\rangle$ from its presence in our expression for the fundamental frequency $\nu_{0}$ in equation Eq.(\ref{fundfreq}).
The differences in crust conductivity that would result from these differing prescriptions are discussed in section \ref{section3}.

In the liquid state, \citet{Potekhin1999} proposed a different approach to deal with mixtures.
In this formalism one writes,
\begin{eqnarray}
\ln\Lambda&\approx& \frac{1}{\langle Z^{2}\rangle}\sum_{j}^{N_{\rm sp}}{x_{j}Z_{j}^{2}\ln\Lambda^{\rm OCP}(\Gamma_{j})}    \label{prescription1999}
\end{eqnarray}
where $\Gamma_{j}=Z_{j}^{5/3}\Gamma_{e}$, $x_j$ refers to the \emph{number fraction} as defined in Eq.(\ref{numberfraction}), and $\ln\Lambda^{\rm OCP}(\Gamma)$ is the Coulomb logarithm (\ref{coul_log_szz}) for an OCP at coupling $\Gamma$.
This prescription for mixtures was also recommended recently in the works of \citet{Cassisi2007} and \citet{Itoh2008} and we corroborate its efficacy in section \ref{mixingrule}.
This presciption (\ref{prescription1999}), when substituted in Eq.(\ref{nusigmakappa}), leads to
\begin{eqnarray}
\nu=\nu_{0}\frac{\langle Z^{2}\ln\Lambda^{\rm OCP}\rangle}{\langle Z^{2}\rangle}\/.
\end{eqnarray}

In sections \ref{mixingrule} and \ref{section3}, we propose and validate with MD simulations a mixing rule for $|ln\Lambda$ that is slightly modified from Eq.(\ref{prescription1999}).
We also show that the impurity parameter formalism is inappropriate at large $Q_{\rm imp}$.

\subsubsection{Discontinuities in transport coefficients at crystallization} \label{discontinuity}

The previous expressions tell us that the conductivities and NPB emissivity depend on the scattering of electrons off the fluctuations $\delta n_{z}({\bf k},t)$ in the ionic charge density about the average value $\langle n_{z}({\bf k})\rangle_{\rm eq}$. However, this average charge density is dependent upon thermodynamic phase of the system:
\begin{itemize}
\item If the system is fluid, the averaged charge density $\langle n_{z}({\bf k})\rangle_{\rm eq}=n_{e}\,\delta({\bf k})$ is uniform;
\item If the system is in a crystalline solid state,  i.e. comprised of a periodic (Bravais) lattice with the same physical unit, electrons are in Bloch states and $\langle n_{z}({\bf k})\rangle=e^{-2W(k)}n_{e}V\sum_{{\bf K}}\delta_{{\bf k},{\bf K}}$ depends on ${\bf k}$, where, assuming $N_{\rm sp}=1$, ${\bf K}$ are reciprocal lattice vectors and $W(k)$ is the Debye-Waller factor: Bloch electrons do not scatter off the static, wavevector-dependent charge fluctuations (Bragg peak).
\end{itemize}

Because of this fundamental difference, the formulae of the previous section predict significant (by a factor of 3-5) discontinuities of the electric and thermal
conductivities at the melting point \citep{Itoh1993}.
Recently, \citet{Baiko1998} argued that this jump in the OCP electronic transport properties upon crystallization is not physical and that the electron transport properties in strongly coupled plasmas should be fairly insensitive to the thermodynamic state of the plasma.
Accordingly, they advance that strongly coupled Coulomb liquids are characterized locally by the same long-range order as present in the crystalline solid and
although the long-range order does not persist forever as in a solid (the liquid state is fluid), electrons keep traveling in those local crystal-like structures without any significant degradation of their mean velocity.
Thus, \citet{Baiko1998} advocate that, in the liquid state, the contribution corresponding to elastic scattering off the incipient crystalline structure must be substracted from the static structure factor.
As intended, this prescription removes the jump in the conductivities at the solid-liquid transition - 
this prescription was later used in a series of papers to redefine practical fitting formulas for the electron conductivities and NPB (e.g, \citet{Cassisi2007} and references therein ; also \citet{Horowitz2008}). 

The prescription of \citet{Baiko1998} contradicts our current understanding of solid and liquid metals.
We instead agree with \citet{Itoh2008} since we believe that the disordered positions and the oscillating and diffusive movement of the ions cannot be ignored in a liquid.

The arguments and conclusions of \citet{Baiko1998} contradict both experimental findings and the contemporary theoretical understanding of liquids: for example, 
it has been known for a long time that the electrical resistance of most metals in the liquid state just above their melting points are about 1.5-2.3 times larger than those of solid metals \citep{CusakEnderby1960}.
Even the simplest metals, whose electronic structure do not change appreciably upon melting, show a discontinuity in transport properties at melting, e.g. for sodium $\sigma_{\rm s}/\sigma_{\rm l}\approx 1.45$ and for aluminum $\sigma_{\rm s}/\sigma_{\rm l}\approx 2.2$.
Interestingly, similar to Baiko et al.'s approach but with the opposite purpose in mind, early theoretical attempts to explain the increase of electrical resistivity upon melting relied heavily on ideas that had proved successful for solid metals. These approaches overlooked the effect of disorder inherent to liquids:  for instance, in the ``quasi-crystalline'' models, the local coordination just above the melting point was treated very similar to that which prevails in the solid phase just below.
It was also believed that, as in a solid metal, the resistivity of liquid metals could be divided into two parts (a) a thermal term, proportional to the mean square amplitude of vibration of the ions, and hence dependent on the temperature $T$, and (b) a residual term, independent of $T$ due to deffects (e.g., vacancies, dislocations).
In 1934, \citet{Mott1934} presented an elegant theory to reconcile these solid-like picture to the increase of electical resistivity on melting in terms of the entropy change on melting per particle $\Delta s=(S_{\rm l}-S_{\rm s})/nV$,
\begin{eqnarray}
\Delta s=1.5 k_{\rm B}\ln \left(\sigma_{\rm s}/\sigma_{\rm l}\right)\/.
\end{eqnarray}
Using the latent heat of fusion to evaluate $\Delta s$, Mott's model yields surprisingly good values of $\sigma_{\rm s}/\sigma_{\rm l}$ ($1.68$ for Sodium instead of $1.45$.)
It was later realized that Mott's assumptions do not hold in the light of other facts.
Hence Mott's theory assumes that all the melting entropy is connected with the thermal agitation of the ions and ignored the configurational entropy arising from disorder in the rest positions about which ions are instantaneously vibrating (the so-called ``cage effect".)
Unfortunately, this argument does not hold:
it is indeed known that for all simple metals, the experimental $\Delta s$ is approximately a universal constant, $\Delta s\approx 0.8\pm 0.1$ (rms), and is chiefly due to the change in configurational entropy \citep{Wallace1997}.
The entropy change due to the thermal agitation, the anharmonicities in ionic motions and the electronic structure are responsible for the small scattering in the universal value.
As illustrated in table \ref{table1}, similar findings apply the OCP over a wide range of electronic screening, namely $\Delta s\approx 0.85$.

Finally, we would like to add another argument in disfavour of Baiko et al.'s prescription.
If crystallites do indeed exist in the liquid phase as assumed by \citet{Baiko1998}, they must be extremely small, smaller than required for a local band structure to be established, because the presence of nuclei would make it impossible for a liquid to supercool. However, all liquid metals can be persuaded to supercool through as much as $20\%$ of their melting temperature \citep{TurnbullCech1950} and MD simulations of the OCP show the same trend \citep{Daligault2007}.

Interestingly, for the multi-component systems found in the crusts of accreting neutron stars, the jump may be much less important than suggested by the calculations of \citet{Itoh1993}. However this is likely not due to the reason proposed by  Baiko et al. but because of the lack of long-range order for those complex mixtures and the high degree of impurity-like disorder that those systems possess.

\subsubsection{Terrestrial liquid metal compared to the neutron star crust plasma}

In this section we continue the comparison of electron-ion scattering as occurring 
in terrestrial liquid metals as opposed to the crust of a neutron star.
To make this comparison as quantitative as possible, we shall discuss the different components of the Coulomb logarithm in Eq.(\ref{CoulombLogarithm}).
While this expression applies to both terrestrial liquid metals and to the crusts of accreting neutron stars (under the assumption of the first Born approximation), there are several interesting differences between both systems
which we enumerate below:

\begin{itemize}

\item[(a)] While in the former the number of species is typically of order unity ($N_{\rm sp}=1$ in simple metals; $N_{\rm sp}=2-3$ in alloys), the number of species in the latter can be several hundred (see section \ref{composition}.)
Nevertheless, as shall see, the charge-charge structure factors $S_{\rm zz}(k)$ is similar in both cases, i.e. characterized by damped oscillations originating from the short-range order of strongly coupled systems.
The heights of the peaks increase with the strength of the coupling parameter $\Gamma_{\rm eff}$ and therefore with depth.

\item[(b)] While in terrestrial liquid metals electrons are non-relativistic, electrons in an accreting neutron star crust are relativistic.
The main effects of relativity are encapsulated in  $\left[1-\frac{x_{\rm r}^{2}}{1+x_{\rm r}^{2}}\left(\frac{k}{2k_{\rm F}}\right)^{2}\right]$, which varies from $1$ when $x_{\rm r}\to 0$ and $1-(k/2k_{\rm F})^{2}$ when $x_{\rm r}\to\infty$ and describes kinematic suppresion of backward scattering of relativistic electrons.
Relavity also modifies the electronic screening of the ions (see Eq.(\ref{epsilon_e}).
The overall effects of relativistic corrections is illustrated in Fig.(\ref{figure3}) and in Fig.(\ref{figure4}).
Figure (\ref{figure3}), shows the integrand of Eq.(\ref{CoulombLogarithm}) for both relativistic and non-relativistic electrons ($x_{\rm r}=0$ in Eq.(\ref{CoulombLogarithm})).
Relativistic effects are noticeable in large momentum transfer collisions (close encounters) and, as a consequence of the suppression of back-scattering,  result in lower scattering cross-sections and consequently higher conductivities.
Fig.(\ref{figure4}) compares the relativistic vs nonrelativistic Coulomb logarithm as a function of depth using the compositional profile of \citet{Gupta2007} discussed in section \ref{composition}.

\item[(c)] While in the neutron star crust, ions are fully stripped and all electrons are delocalized electrons and participate in the electronic conduction, $n_{e}=\langle Z\rangle n$ with $\langle Z\rangle\sim 40$, most electrons in a liquid metal are bound to the nuclei, $n_{e}=\langle Z_{\rm eff}\rangle n$ where $Z_{\rm eff}$ is the effective charge of the ions and $Z_{\rm eff}=1-3$.
As a consequence, in Eq.(\ref{CoulombLogarithm}), $v_{ie}(k)$ must in principle be replaced by a pseudopotential that mimics the effect of bound electrons on free-electron ion scattering (this is a major difficulty in condensed matter physics). In contrast, for neutron star crusts, $v_{ie}$ is known: it is the pure Coulomb potential, slightly corrected with a form factor to describe the finite size of the nuclei and very close electron-ion collisions.

\item[(d)] A final effect of major importance arises from the magnitude of the electron Fermi wavevector $k_{\rm F}=(3\pi^{2}n_{e})^{1/3}$, which increases like $Z_{\rm eff}^{1/3}$ in terrestrial liquid metals as opposed to  $\langle Z\rangle^{1/3}$ in the neutron star crust.
Therefore the ranges of integration $0\leq k\leq 2k_{\rm F}$ in Eq.(\ref{CoulombLogarithm}) greatly vary between the two cases.
The situation is illustrated in Fig.(\ref{figure3}):
the limit of integration $2k_{\rm F}$ lies just to the left of the main peak in $S(k)$ for monovalent liquid metals, just to its right for divalent ones, and progressively further to the right for metals which have a valency greater that two and three.
In neutron stars, $2k_{\rm F}$ can be quite large (and increases with depth), i.e. large momentum transfer collisions (close encounters) become much more important in determining the electron-ion scattering cross-section.

\end{itemize}

\section{Molecular Dynamics (MD) Simulations of the neutron star crust} \label{sectionMD}

In this section, we discuss the MD simulations we performed in order to calculate the structure factors $S_{\rm zz}(k)$ of MCPs representative of neutron star crust and to validate the linear mixing rule discussed in section \ref{mixingrule}.

\subsection{Input : a representative crust composition profile} \label{composition}

To base our calculations of crustal $\kappa$ and $\epsilon_{\nu}$ on a realistic pre-neutron-drip crust compositional profile we have used the compositions of \citet{Gupta2007}. Our choice was determined by a number of factors. First, the diversity of species in the starting composition (XRB ashes from \citet{SchatzXRB} is characterized by the ``impurity parameter"  which has a high value $Q_{\rm imp} \sim100$ , an ideal test-bed for validation of a mixing rule in an MCP. In the outer crust, prior to the onset of neutron reactions, electron captures preserve the high impurity. Only beyond neutron-drip in the inner crust, as demonstrated in the SEC-nucleosynthesis process described in \citet{Gupta2008}, do electron-capture-delayed-neutron emissions rearrange abundances between very diverse mass chains. Finally, the accretion rate closely match those expected of X-ray superburst progenitors ($\sim 0.1 \dot{M}_{\rm Edd} \approx 10^{17}$ g s$^{-1}$) and therefore we intend the numbers we compute to have practical relevance to the neutron star crust modeling and X-ray observational communities. 
  
The mixtures we have simulated discard the less abundant species which would have had less than $10$ particles in the simulations.
Details for each of the compostions used at depth are given in table \ref{table2} and the spread in $Z$ is illustrated in Fig.(\ref{figure5}) for two compositions.

We now use the composition profiles at various neutron star crust depths to validate the linear mixing rule. We adopt a fiducial temperature $T=0.5$ GK, and vary the density as in Table \ref{table2}. Thus, changes in $\Gamma_{\rm eff}$ with changing crust depth arise only from composition evolution and a decrease in inter-ionic spacing . At each depth a large number of mass chains $A\approx20-100$ enter the composition as expected from X-ray burst ashes evolving in the outer crust at depths shallower than the neutron-drip point.  

\subsection{The MD Procedure}

We employ molecular dynamics (MD) simulations to calculate the structure factor (\ref{Sqq}) at various depths of a neutron star crust.
In order to deal with the impure multi-component ionic mixtures typical of neutron star crusts, we have developed a code that allows for the simulation of large charged systems (tens to a few hundred nuclear species) over the long time scales required for the equilibration of such impure MCPs.
The code is based on a parallelized implementation of the Particle-Particle-Particle-Mesh ($\rm P^{3}M$) algorithm with periodic boundary conditions; high resolution for individual encounters is combined with rapid, mesh-based, long range force calculations.
It simulates the classical dynamics of a mixture of nuclei described as point particles interacting through a screened Coulomb interaction.
Thus the interaction potential between two nuclei of charge $Z_{i}$ and $Z_{j}$ is modeled by the screened Coulomb (Yukawa) potential,
\begin{eqnarray}
v_{ij}(r)=\frac{Z_{i}Z_{j}e^{2}}{r}e^{-k_{\rm sc}r}\/, \label{screenedpotential}
\end{eqnarray}
where $k_{\rm sc}$ is the inverse screening length.
The $\rm P^{3}M$ algorithm can handle any value $k_{\rm sc}\ge 0$ to very high accuracy.
In practice, we use the inverse of the relativistic Thomas-Fermi inverse screening length $k_{\rm rTF}$ for $k_{\rm sc}$ given by $k_{\rm rTF}a_{e}=0.185\/(1+x_{r}^{-2})^{1/4}$.

Several technical details of the MD calculations (simulation length, number of particles, time steps, etc.) are collected in appendix \ref{appendix1}.
Here we briefly outline the procedure used to calculate $S_{\rm zz}(k)$.
Given a composition at a certain depth, the particles are initially randomly distributed in a cubic box of volume $V=L^{3}$ with a Maxwellian velocity distribution at the prescribed temperature.
After a long equilibration phase, a simulation is performed in which, at each time step $n$, the charge density $n_{Z}({\bf k};n)$ as in Eq.(\ref{rhoz}) is calculated for values of ${\bf k}$ commensurate with the simulation box size $L$, namely
\begin{eqnarray}
{\bf k}=\frac{2\pi}{L}(n_{x},n_{y},n_{z})\quad,\quad ||{\bf k}||\leq 2k_{\rm F}\, \label{kMD}
\end{eqnarray}
where $n_{x,y,z}$ are natural integers.
The charge-charge structure factor (\ref{Sqq}) is calculated at the end of the run using
\begin{eqnarray}
S_{\rm zz}(k)=\frac{1}{N_{run}}\sum_{n=1}^{N_{run}}{\sum_{{\bf k}\,,\,||{\bf k}||=k}{\frac{1}{N_{k}}|n_{z}({\bf k},n)|^{2}}}\/. \label{SzzMD}
\end{eqnarray}

\subsection{Molecular dynamics structure factors}

MD results for the structure factor $S_{\rm zz}(k)$ and corresponding to the crustal compositions of \citet{Gupta2007} listed in table (\ref{table2}), all at temperature $T=0.5$ GK, are shown in Figs.(\ref{figure6}-\ref{figure8}) over the range  $0\leq k\leq 2k_{\rm F}$.
Additional data obtained from the MD simulations to test lower temperatures near the photosphere-ocean boundary and potentially high temperatures at different depths are provided in the Appendix \ref{appendixSk}.

The shape of the charge-charge structure factor $S_{\rm zz}(k)$ is typical of the structure factor of liquids.
The first peak, which systematically occurs at $k/2k_{\rm F}\approx 0.33$ and which has a height increasing with the effective Coulomb coupling $\Gamma_{\rm eff}$, reflects the existence of a dominant short-range order of the particles in real space.
This is illustrated in Fig.(\ref{figure14}) that shows different pair-distribution functions $g_{ab}(r)$ for several pairs of species $a$ and $b$.
The sharp decrease in the pair-distribution functions at small separation, originating from the repulsion between like charges, is responsible for the subsequent maxima and minima of $S_{\rm zz}(k)$, whose oscillation is strongly damped as $k$ increases.
Eventually, at large $k$, $S_{\rm zz}(k)$ approaches unity, in accordance with the normalization in Eq(\ref{Sqq}).
At small $k$, $S_{\rm zz}(k)$ probes the long-wavelength static fluctuations in the charge density and approaches zero (as $k^{2}$) in the $k\to 0$ limit as a consequence of charge neutrality and perfect screening.

All the simulations at density larger than $\sim 10^{10}$ $\rm g.cm^{-3}$ (compositions \#9-17 of table \ref{table2}) and characterized by $\Gamma_{\rm eff}>260$ show pronounced short-range structure as seen in the shoulder of the second peak of $S_{\rm zz}(k)$.
Nevertheless, within the timescale of the simulations, we did not witness any clear-cut phase separation or crystallization.

\section{Validation of the Linear Mixing rule and Practical Fit} \label{mixingrule}

\subsection{Validation of the LMR}

A common approximation for the excess (non-ideal) energy $U_{\rm ex}$ and free energy $F_{\rm ex}$ of strongly coupled ion mixtures is the (empirical)  Linear Mixing Rule (LMR).
In terms of the energy per particle and per unit of $k_{B}T$, $u_{\rm ex}=U_{\rm ex}/Nk_{B}T$ and $f_{\rm ex}=F_{\rm ex}/Nk_{B}T$, the LMR claims that
\begin{eqnarray}
u_{\rm ex}(\Gamma_{e},\overrightarrow{\mathbf{x}})&\approx& u_{\rm ex}^{\rm LMR}(\Gamma_{e},\overrightarrow{\mathbf{x}})\nonumber\\
&=&\sum_{j=1}^{N_{\rm sp}}{x_{j}u_{\rm ex}^{\rm OCP}(\Gamma_{j})} \label{LMFforuex}\\ 
f_{\rm ex}(\Gamma_{e},\overrightarrow{\mathbf{x}})&\approx& f_{\rm ex}^{\rm LMR}(\Gamma_{e},\overrightarrow{\mathbf{x}})\nonumber\\
&=&\sum_{j=1}^{N_{\rm sp}}{x_{j}f_{\rm ex}^{\rm OCP}(\Gamma_{j})} \label{LMFforfex}
\end{eqnarray}
where $\Gamma_{j}=Z_{j}^{5/3}\Gamma_{e}$, and $u_{\rm ex}^{\rm LMR}$ and $f_{\rm ex}^{\rm LMR}$ are the excess free energy of an OCP at coupling $\Gamma_{j}$  and $\overrightarrow{\mathbf{x}}=\{x_{i}\}_{i=1}^{i=N_{\rm sp}}$ is the composition vector of species number fractions (see Eq.(\ref{numberfraction})) for the MCP.

To the best of our knowledge, the mixing rules (\ref{LMFforuex}) and (\ref{LMFforfex}) were obtained empirically and there is no rigorous ``derivation'' of them.
Equation (\ref{LMFforfex}) was shown to be very accurate for binary mixtures with both rigid \citep{DeWitt2003} and polarizable background electrons \citep{Chabrier1990}.
Recently, \citet{PotekhinChabrierRogers} proposed to extend the LMR to calculation of the equation of state (EOS) of multi-component mixtures.
The results derived in the following validate this recent prescription of \citet{PotekhinChabrierRogers}.

The LMRs (\ref{LMFforuex}) and (\ref{LMFforfex}) suggest a more fundamental, microscopic mixing rule that directly involves the structure factors.
To this end, we recall that the excess energy of a screened MCP (i.e. with ions interaction via the screened Coulomb potential (\ref{screenedpotential})) can be obtained from the charge-charge structure factor $S_{\rm zz}(k)$ using \citep{Rosenfeld1993}
\begin{eqnarray} 
\lefteqn{u_{\rm ex}(\Gamma_{e},\overrightarrow{\mathbf{x}})=}&& \label{excess energy}\\
&&\frac{\Gamma_{e}}{\pi}\frac{\langle Z^{2}\rangle}{\langle Z\rangle^{1/3}}\int_{0}^{\infty}{dk \frac{k^{2}}{k^{2}+k_{\rm sc}^{2}}{\left[S_{\rm zz}(k)-1\right]}}-\frac{k_{\rm sc}\langle Z^{2}\rangle}{2}\/.\nonumber
\end{eqnarray}
The exact equation (\ref{excess energy}) together with the accurate approximation Eq.(\ref{LMFforfex}) suggest the following
\emph{Microscopic} Linear Mixing Rule (MLMR) approximation for the MCP structure factor $S_{\rm zz}(k)$,
\begin{eqnarray}\label{Sqqmixing}
S_{\rm zz}(k;\Gamma_{e},\overrightarrow{\mathbf{x}})&\approx& S_{\rm zz}^{\rm MLMR}(k;\Gamma_{e},\overrightarrow{\mathbf{x}})\label{SofkLMR}\\
&=&\frac{\langle Z\rangle^{1/3}}{\langle Z^{2}\rangle}\sum_{j=1}^{N_{\rm sp}}{x_{j}Z_{j}^{5/3}S_{j}^{\rm OCP}(k;\Gamma_{j})}\nonumber
\end{eqnarray}
where $S_{j}^{\rm OCP}(k;\Gamma)$ is the structure factor of the OCP comprised of species $j$ at coupling $\Gamma$.
Indeed, introducing Eq.(\ref{SofkLMR}) in Eq.(\ref{excess energy}), we obtain the LMR (\ref{LMFforuex}) and (\ref{LMFforfex}).
By introducing Eq.(\ref{SofkLMR}) in Eq.(\ref{CoulombLogarithm}), we obtain the LMR for the Coulomb logarithm,
\begin{eqnarray}
\ln\Lambda&\approx& \frac{\langle Z\rangle^{1/3}}{\langle Z^{2}\rangle}\sum_{j}^{N_{\rm sp}}{x_{j}Z_{j}^{5/3}\ln\Lambda^{\rm OCP}(\Gamma_{j},Z_{j},x_{\rm r})}\/,    \label{prescription2009}
\end{eqnarray}
where $\ln\Lambda^{\rm OCP}(\Gamma_{j},Z_{j},x_{\rm r})$ is the Coulomb logarithm for an OCP of charge $Z_{j}$ at the coupling $\Gamma_{j}$ and relativistic parameter $x_{\rm r}$ (the explicit dependance on the charge $Z_{j}$ comes from the upper limit $2k_{F}=2(3\pi^{2}n_{e})^{1/3}=2(3\pi^{2}Z_{j}n_{j})^{1/3}$ of the integral in Eq.(\ref{CoulombLogarithm})).

\begin{figure} 
\includegraphics[scale=.50]{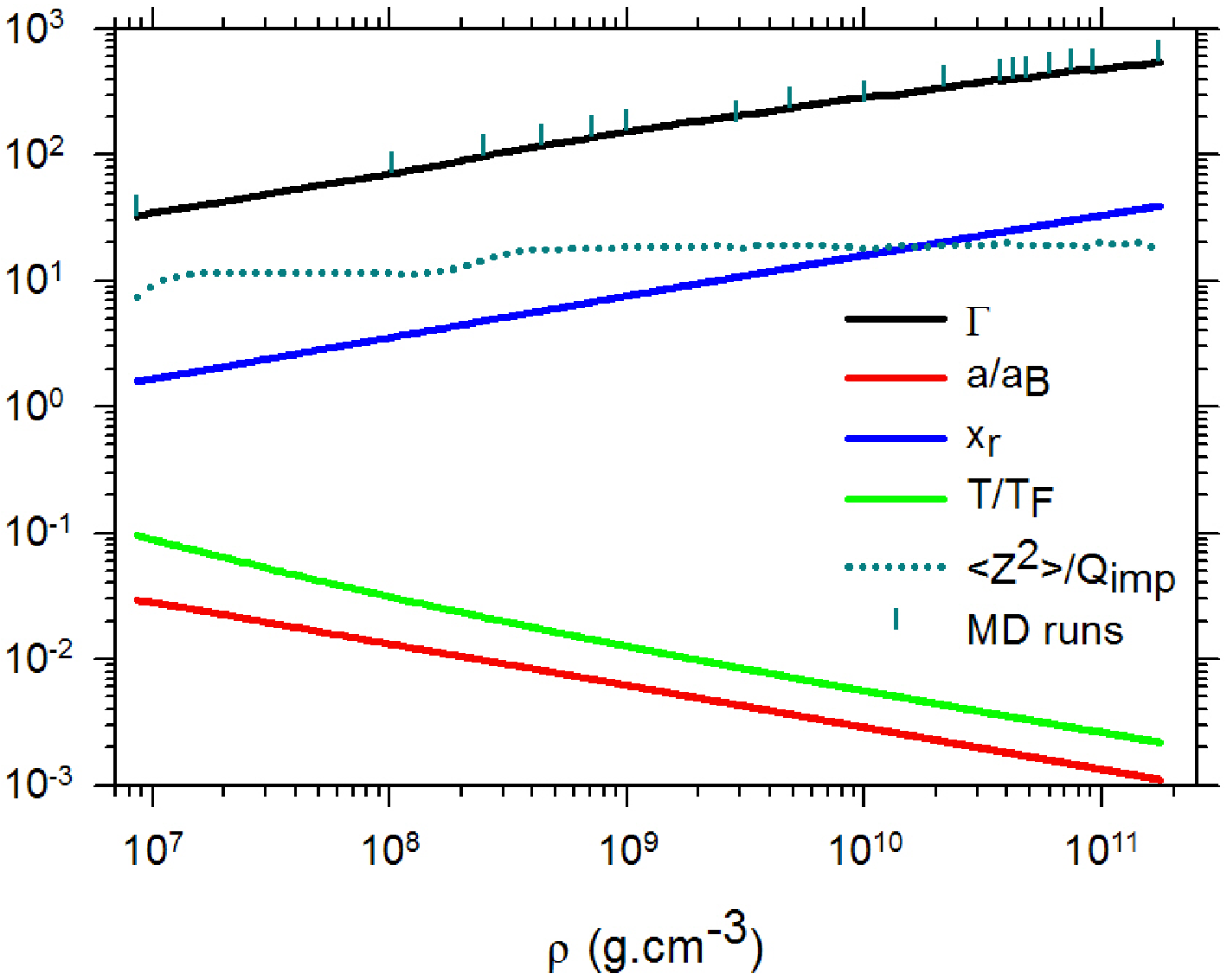}
\caption{Evolution with depth of dimensionless parameters obtained using the crustal composition profile of \citet{Gupta2007}.
$\Gamma_{\rm eff}$ is the effective ion coupling parameter defined in Eq.(\ref{Gamma}), $a/a_{\rm B}$ is the ratio of the mean interionic distance to the electronic Bohr radius, $x_{\rm r}$ is the relativity parameter, $\Theta=T/T_{\rm F}$ is the degeneracy parameter.
The vertical dashes indicate the compositions studied with MD and listed in table \ref{table2}.
The dots show the ratio between the impurity parameter $Q_{\rm imp}$ defined in Eq.(\ref{Qimp}) and $\langle Z^{2}\rangle$.
As discussed in section \ref{section3}, $\langle Z^{2}\rangle$ is less sensitive than $Q_{\rm imp}$ to composition changes at densities $\sim 10^{7}$, $2\times 10^{8}$, $10^{10}$ and $10^{11}$ $\rm g.cm^{-3}$.
\label{figure1}}
\end{figure}

\begin{figure} 
\includegraphics[scale=.50]{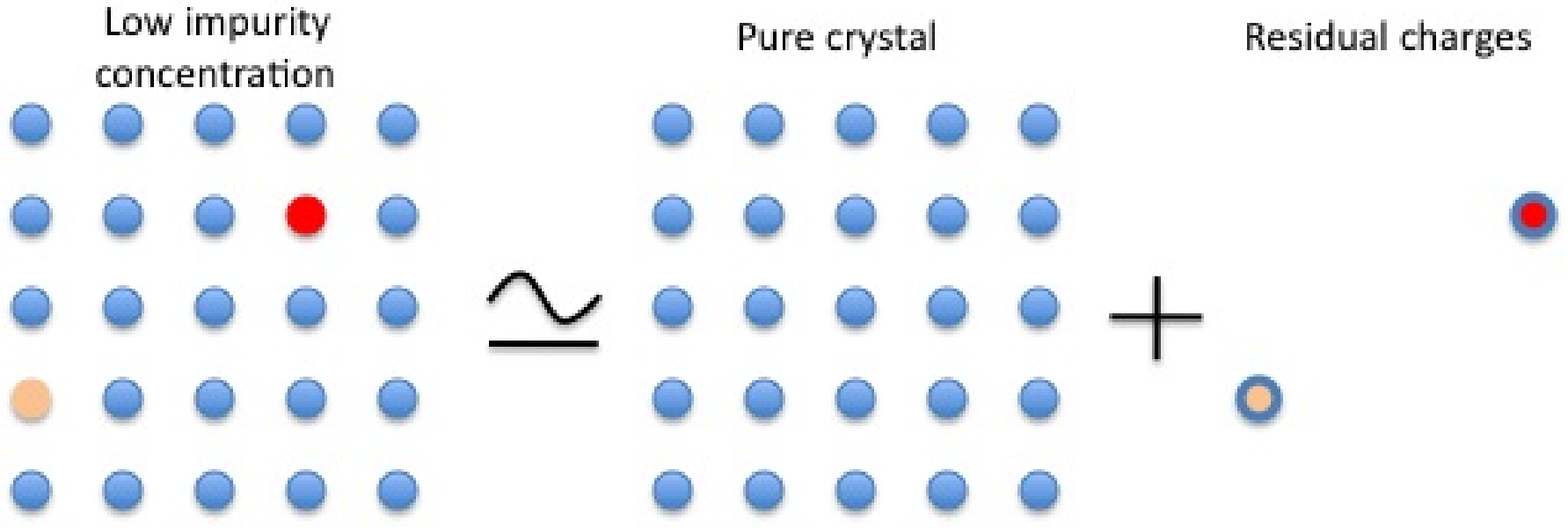}
\includegraphics[scale=.50]{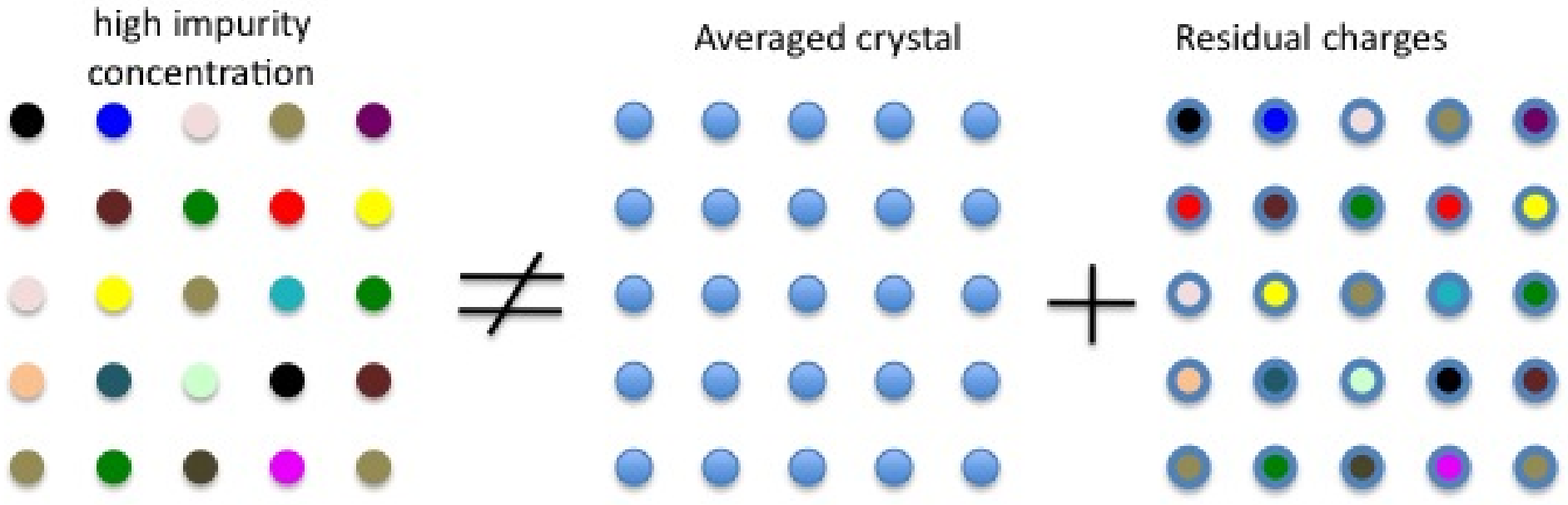}
\caption{The impact parameter formalism illustrated.
(Top panel) A perfect crystal with a low concentration of impurities substituted at lattice sites can be approximated as being made of two uncorrelated parts: the perfect lattice together with the residual charge.
The total electron-ion scattering cross-section is the sum of the Bloch electron-lattice scattering cross-section plus the sum of uncorrelated binary electron-residual impurity collsional cross-section.
(Lower panel) As the concentration of impurities increases, this splitting is less and less justifiable.
\label{figure2}}
\end{figure}

\begin{figure} 
\includegraphics[scale=.50]{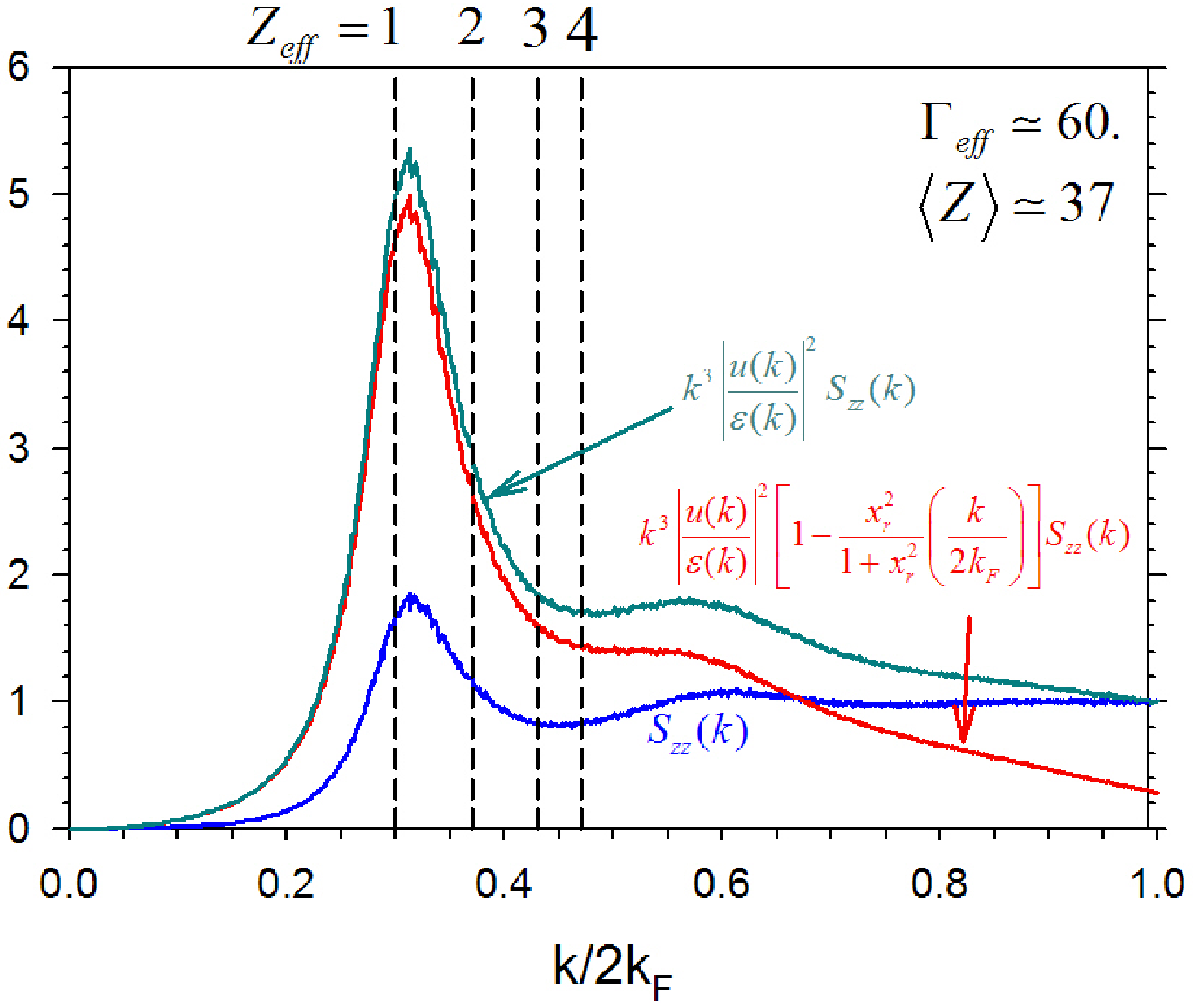}
\includegraphics[scale=.50]{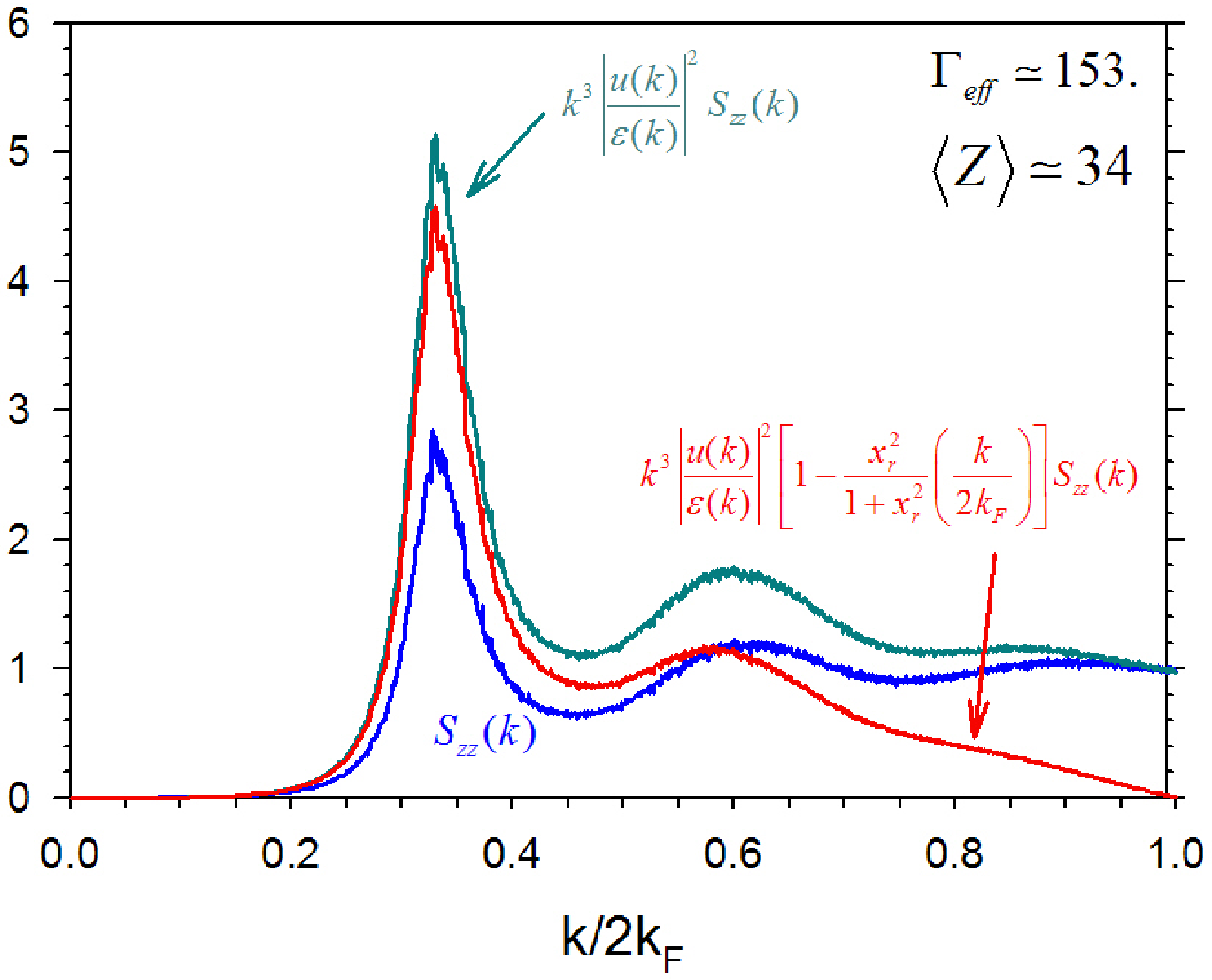}
\caption{(color) The structure factor $S_{\rm zz}(k)$ and the corresponding integrand for the Coulomb logarithm Eq.(\ref{CoulombLogarithm}) with (red line) and without (green line) the relativistic correction.
The upper panel corresponds to the composition \#1 of table \ref{table2} at $T=0,27$ GK ($\Gamma_{\rm eff}=60$ and $\langle Z\rangle=37$).
The lower panel was obtained with compostion \#9 of table \ref{table2} at $T=0.8$ GK ($\Gamma_{\rm eff}=153$ and $\langle Z\rangle=34$).
The vertical dashed lines mark the limit of integration $2k_{\rm F}$ in the Coulomb logarithm (\ref{CoulombLogarithm}) for liquid metals with effective ionic charges $Z_{\rm eff}=1,2,3$ and $4$.
Its position relative to the main peak in $S_{\rm zz}(k)$ clearly depends on the valency of the metal ion.
In the crust of a neutron star, $\langle Z\rangle >> Z_{\rm eff}$, and consequently the range of integration encompasses all first peaks in $S_{\rm zz}(k)$.
Physically, in neutron star crusts, large momentum transfers (close encounters) significantly contribute to the Coulomb logarithm.
\label{figure3}
}
\end{figure}

\begin{figure} 
\includegraphics[scale=.50]{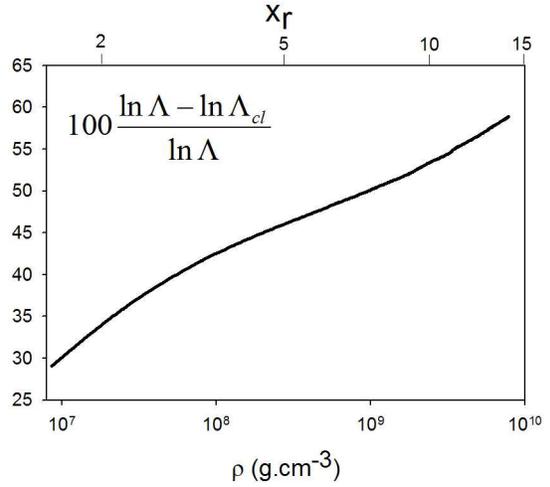}
\caption{(color) Contribution of the relativistic correction to the Coulomb logarithm (\ref{CoulombLogarithm}) in the region $10^{7}\leq\rho\leq 10^{10}$ $\rm g.cm^{-3}$. $\ln\Lambda_{cl}$ corresponds to the nonrelativistic limit of Eq.(\ref{CoulombLogarithm}) obtained by setting the bracketed term $1-\frac{x_{\rm r}^{2}}{1+x_{\rm r}^{2}}\left(\frac{k}{2k_{\rm F}}\right)^{2}$ to $1$.
\label{figure4}
} 
\end{figure}

\begin{figure} 
\includegraphics[scale=.50,angle=-90]{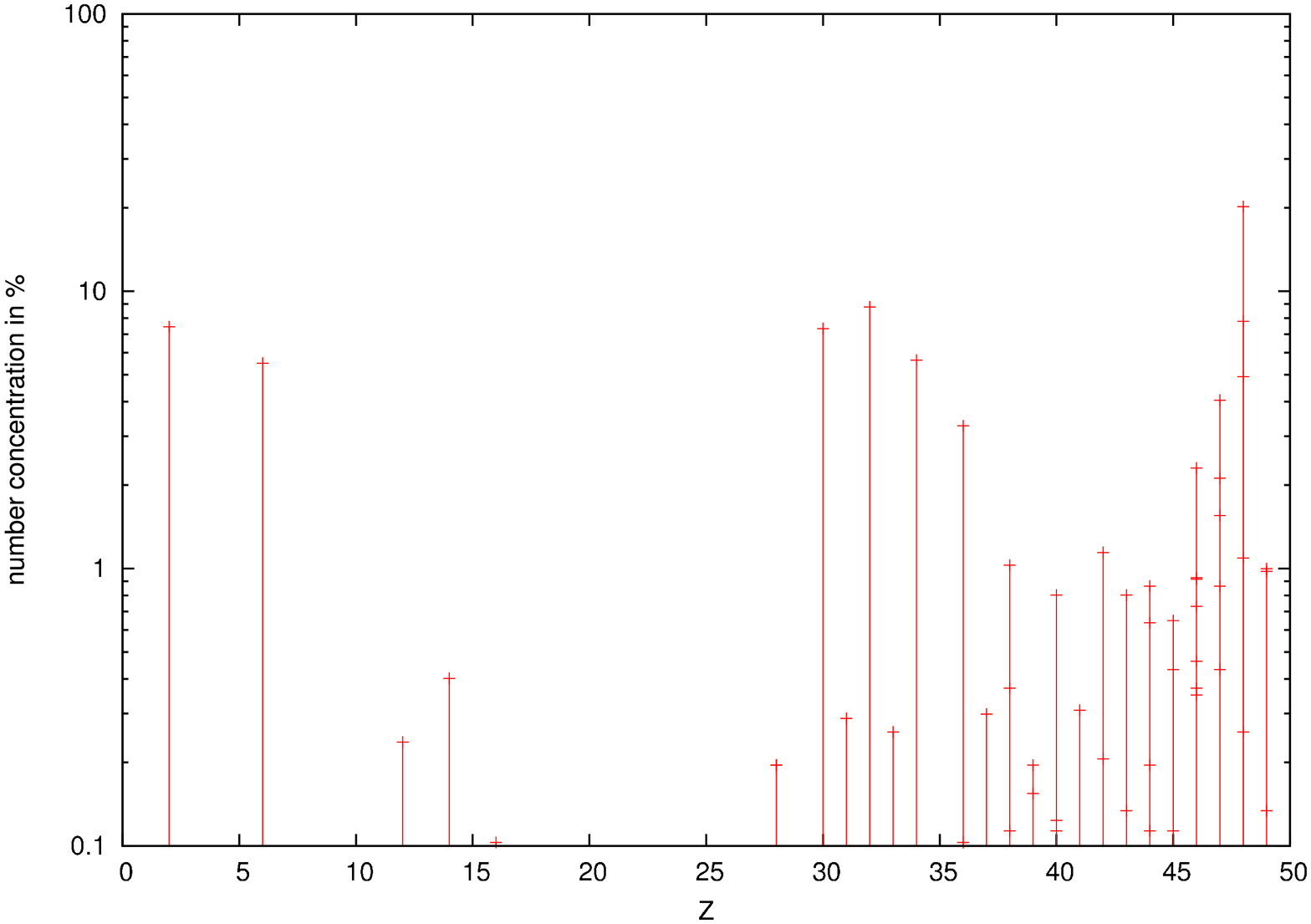}\\
\includegraphics[scale=.50,angle=-90]{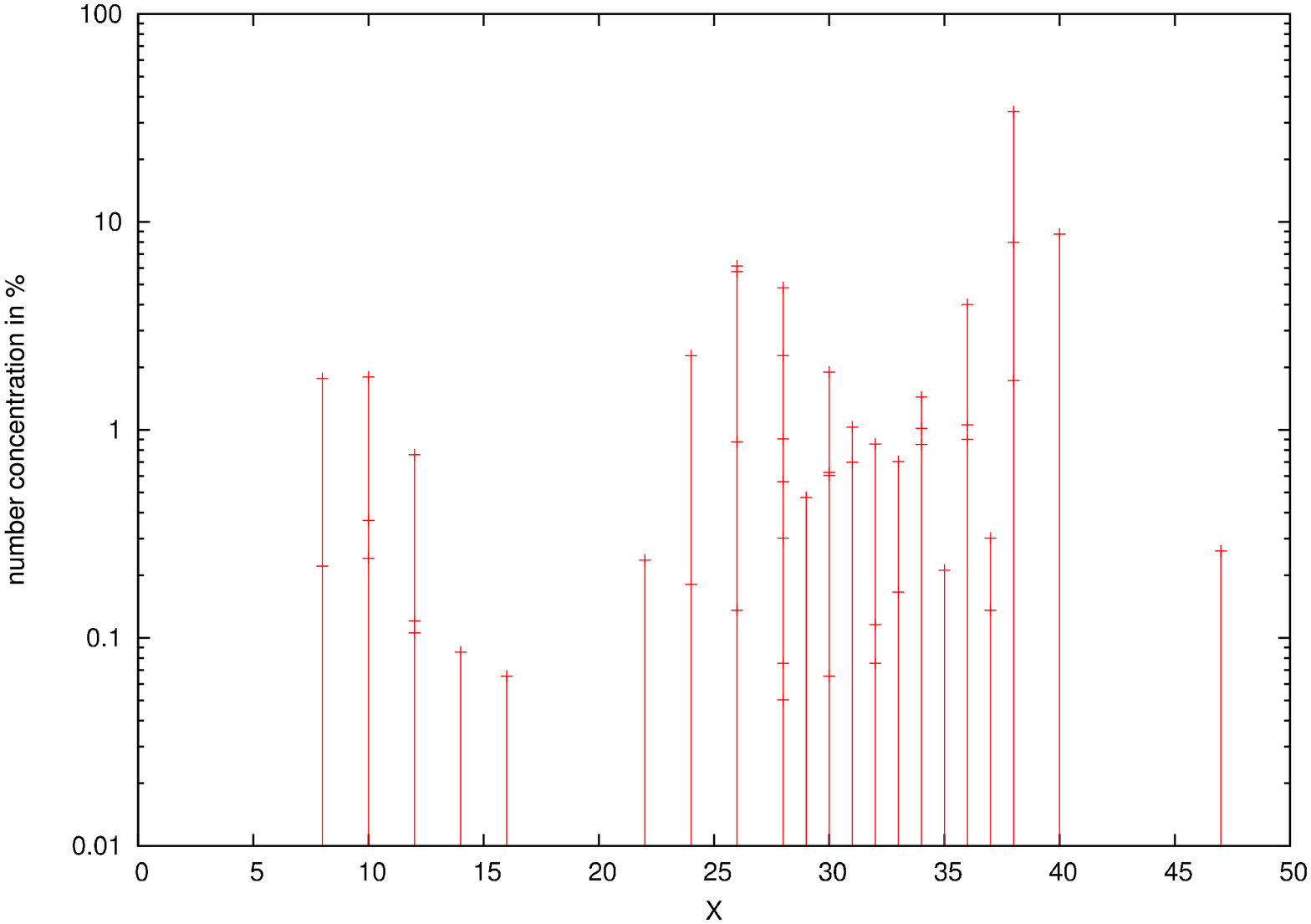}
\caption{Composition profile at two crustal depths corresponding to mixture \#1 (upper panel) and \#12 (lower panel) of table \ref{table2}.
The figure shows the number concentration as a percentage as a function of the nuclear charge $Z$.
At a given $Z$, the horizontal tics indicate different nuclear masses $A$.
\label{figure5}
}
\end{figure}

\begin{figure} 
\includegraphics[scale=.50,angle=0]{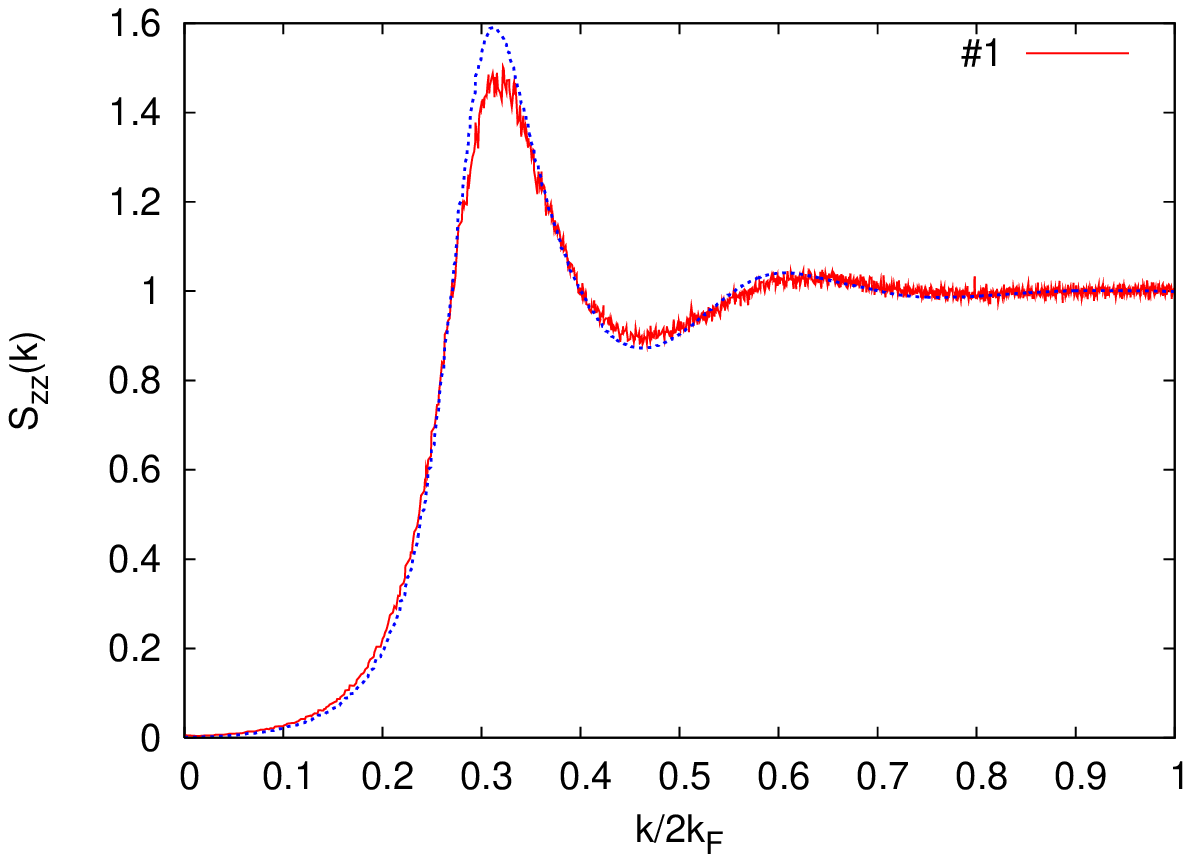}
\includegraphics[scale=.50,angle=0]{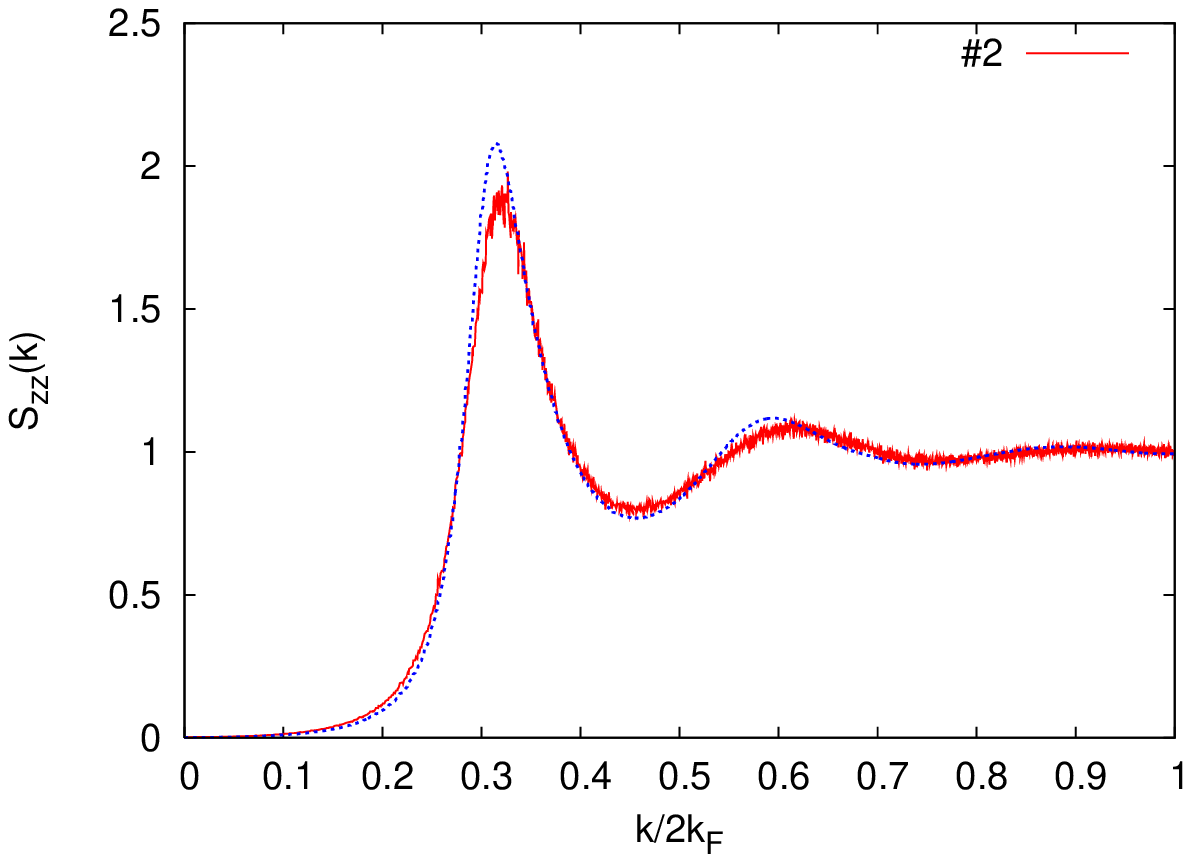}
\includegraphics[scale=.50,angle=0]{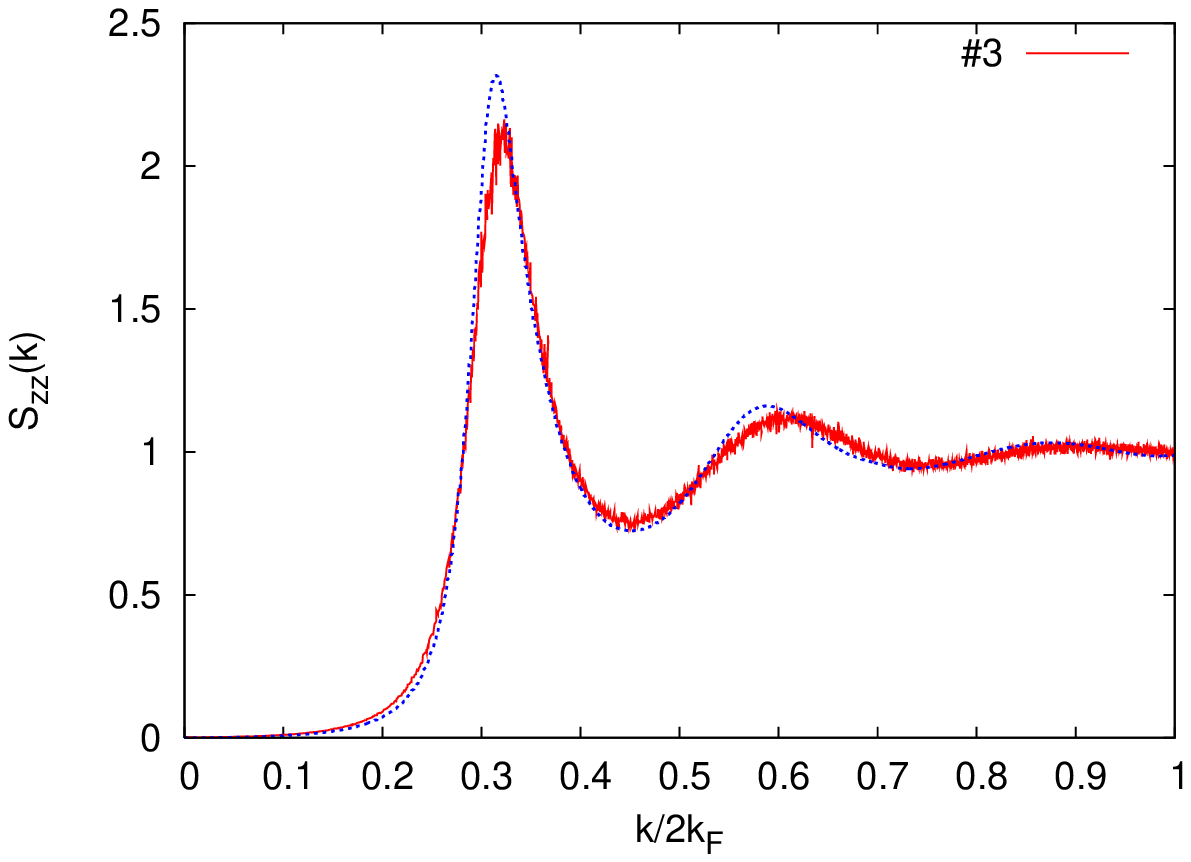}
\includegraphics[scale=.50,angle=0]{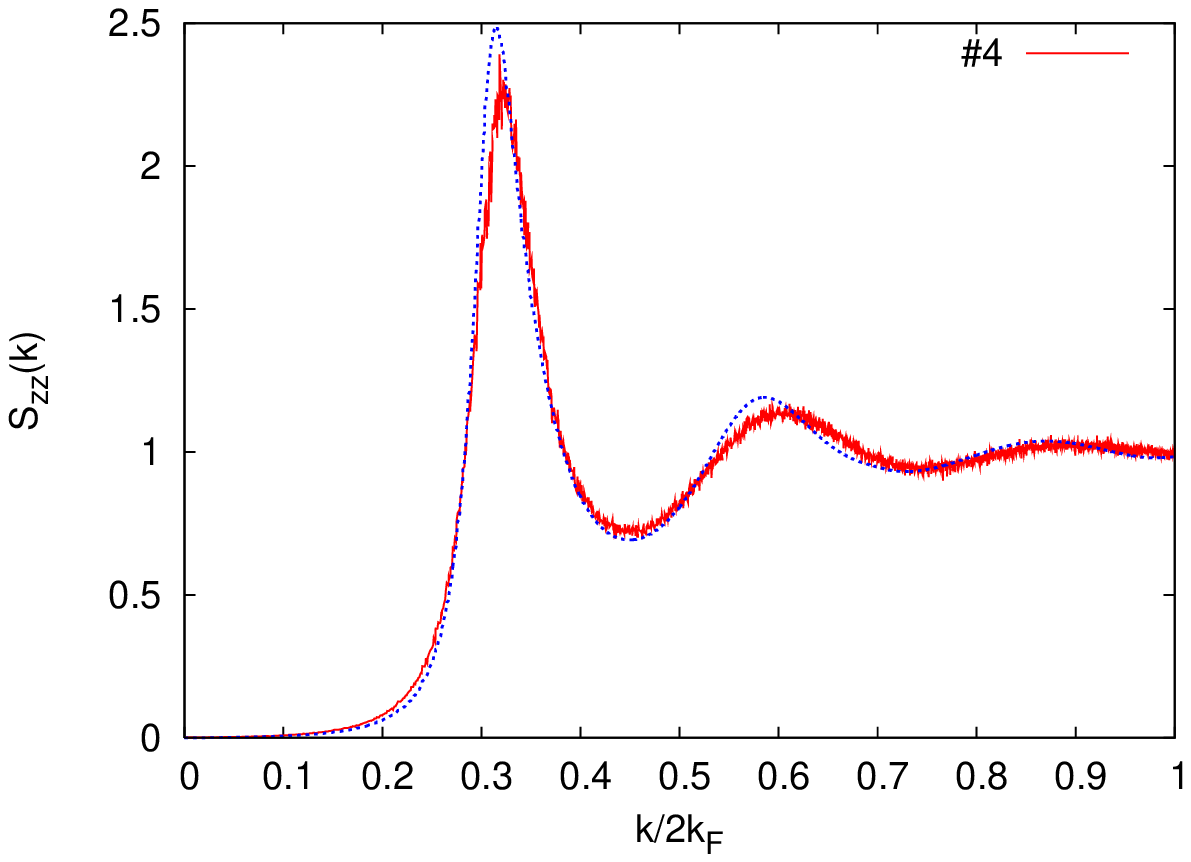}
\includegraphics[scale=.50,angle=0]{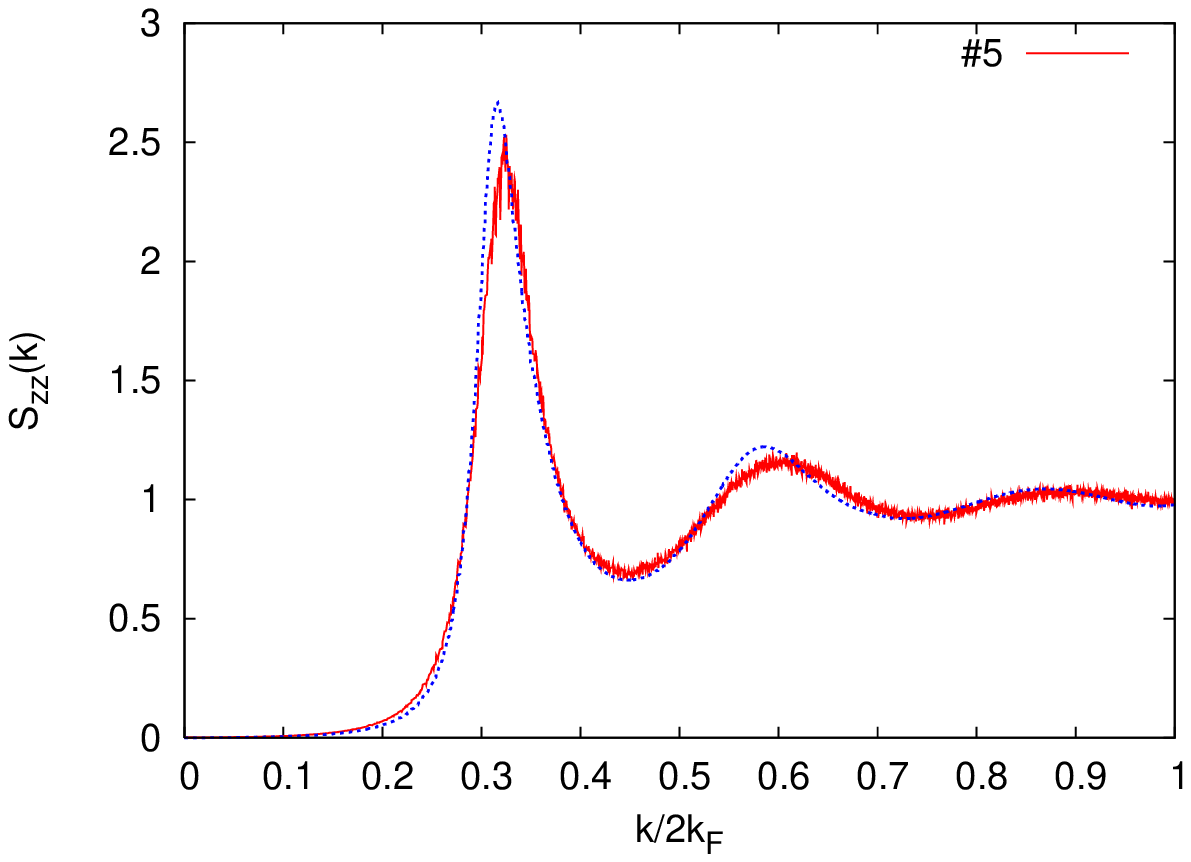}
\caption{(color) The structure factor $S_{\rm zz}(k)$ obtained with MD simulations (red lines) and
with the mixing rule formula (\ref{SofkLMR}) (blue lines) for the mixtures \#1-5 defined in table \ref{table2}.
\label{figure6}
}
\end{figure}

\begin{figure} 
\includegraphics[scale=.50,angle=0]{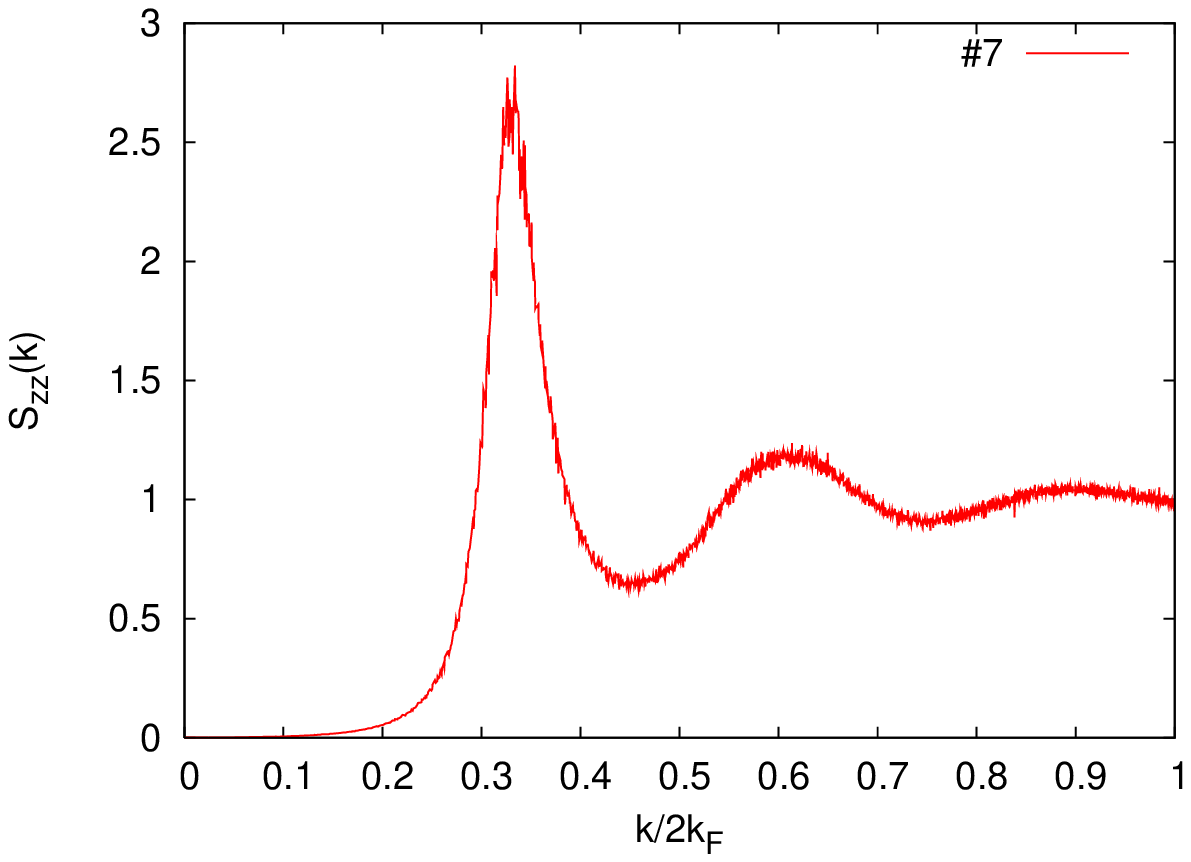}
\includegraphics[scale=.50,angle=0]{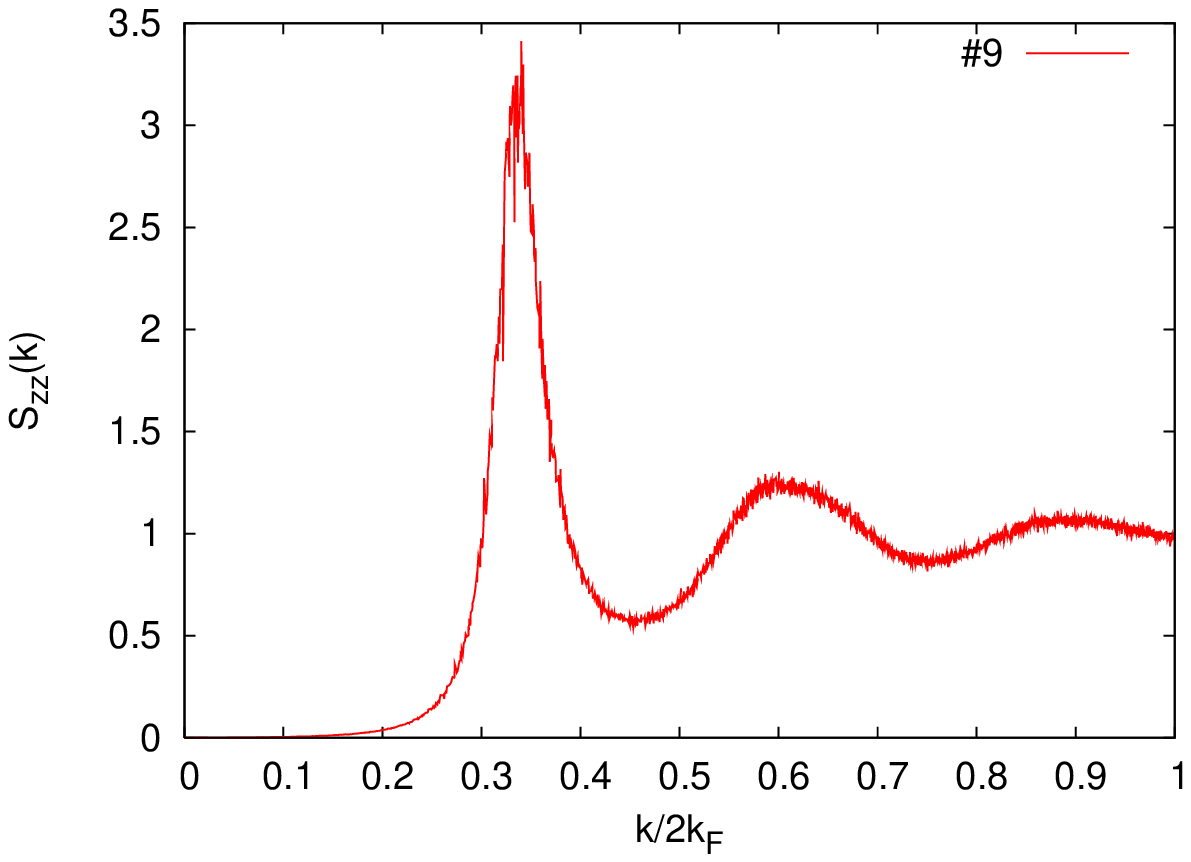}\\
\includegraphics[scale=.50,angle=0]{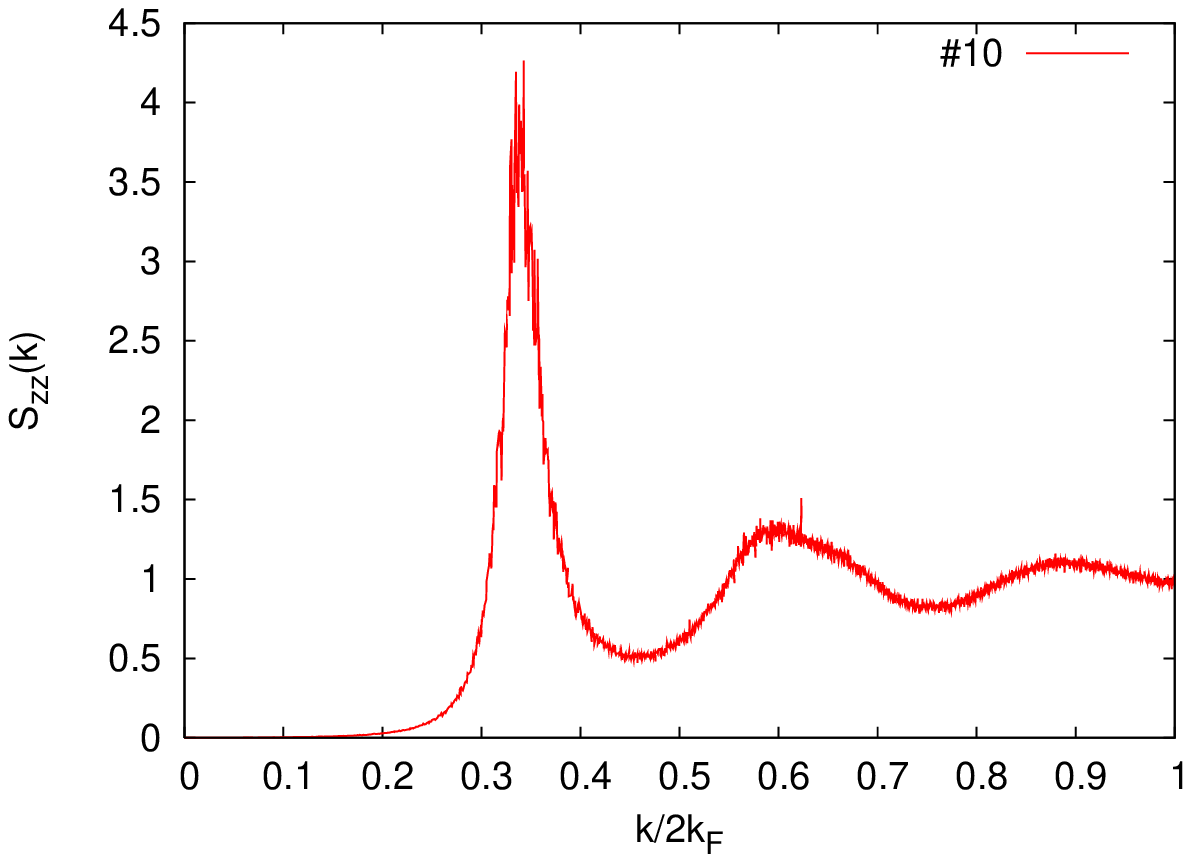}
\includegraphics[scale=.50,angle=0]{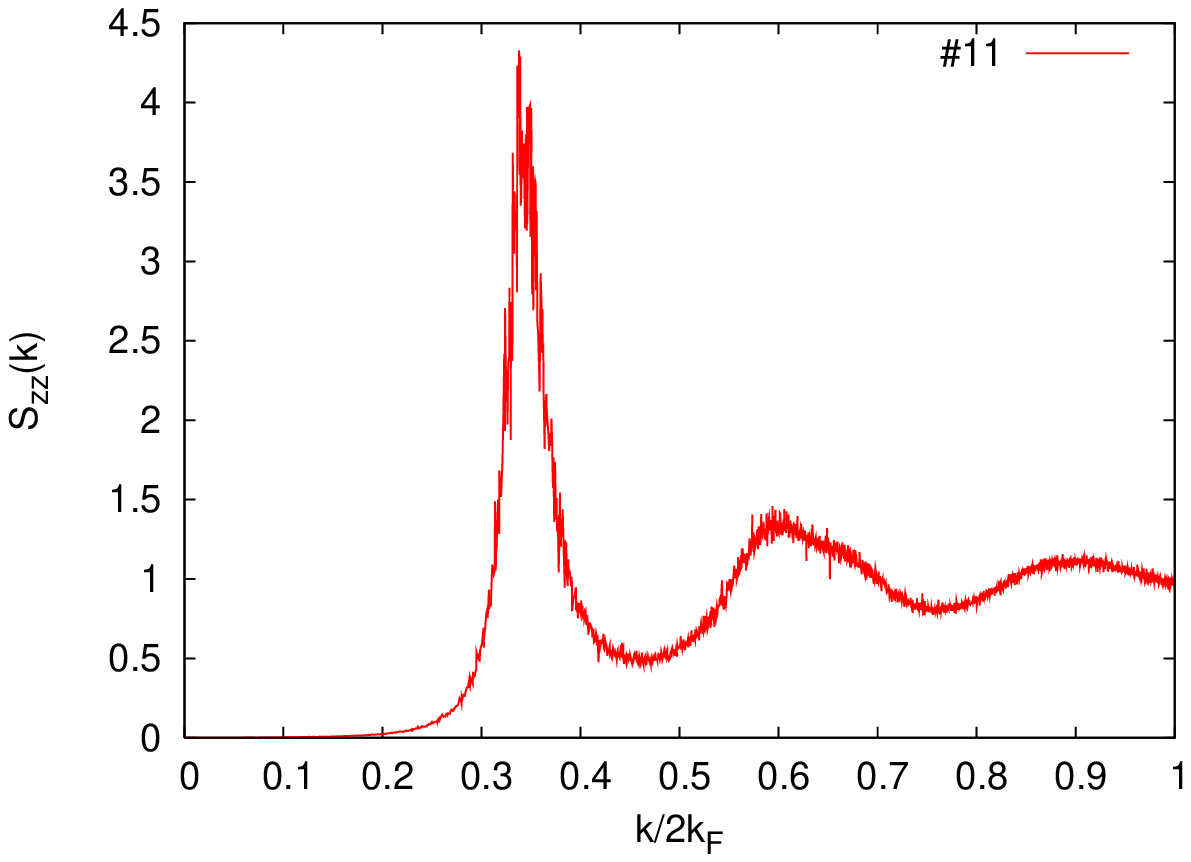}
\caption{(color) The structure factor $S_{\rm zz}(k)$ obtained with MD simulations for the mixtures \#7, \#9, \#10, and \#11 defined in table \ref{table2}.
\label{figure7}
}
\end{figure}

\begin{figure} 
\includegraphics[scale=.50,angle=0]{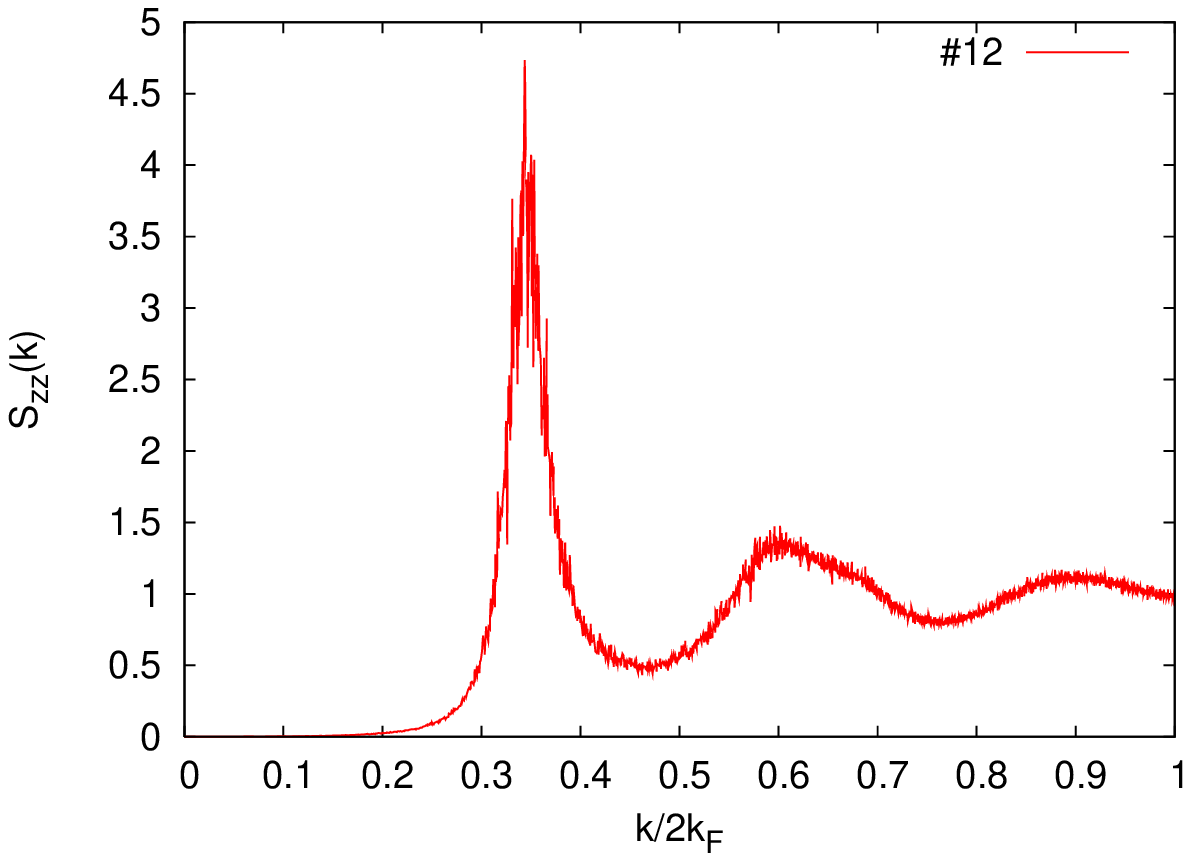}
\includegraphics[scale=.50,angle=0]{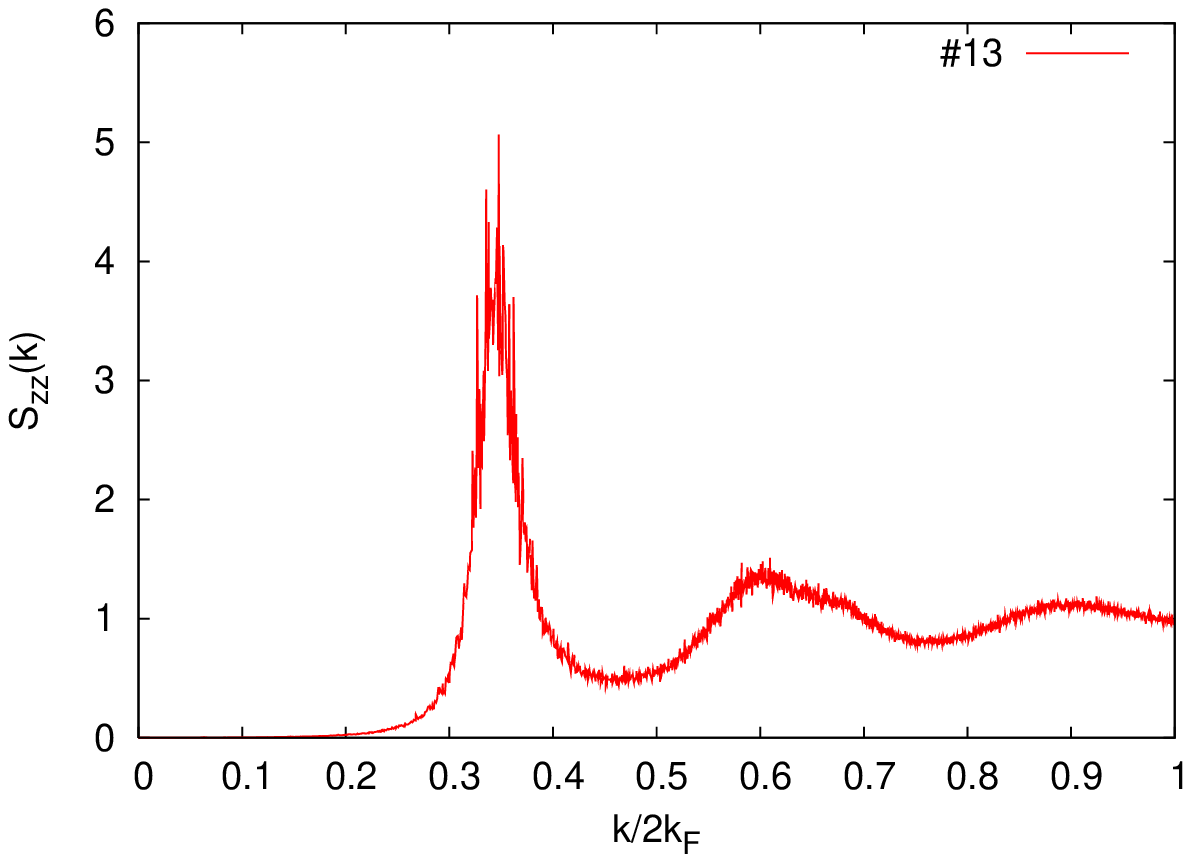}\\
\includegraphics[scale=.50,angle=0]{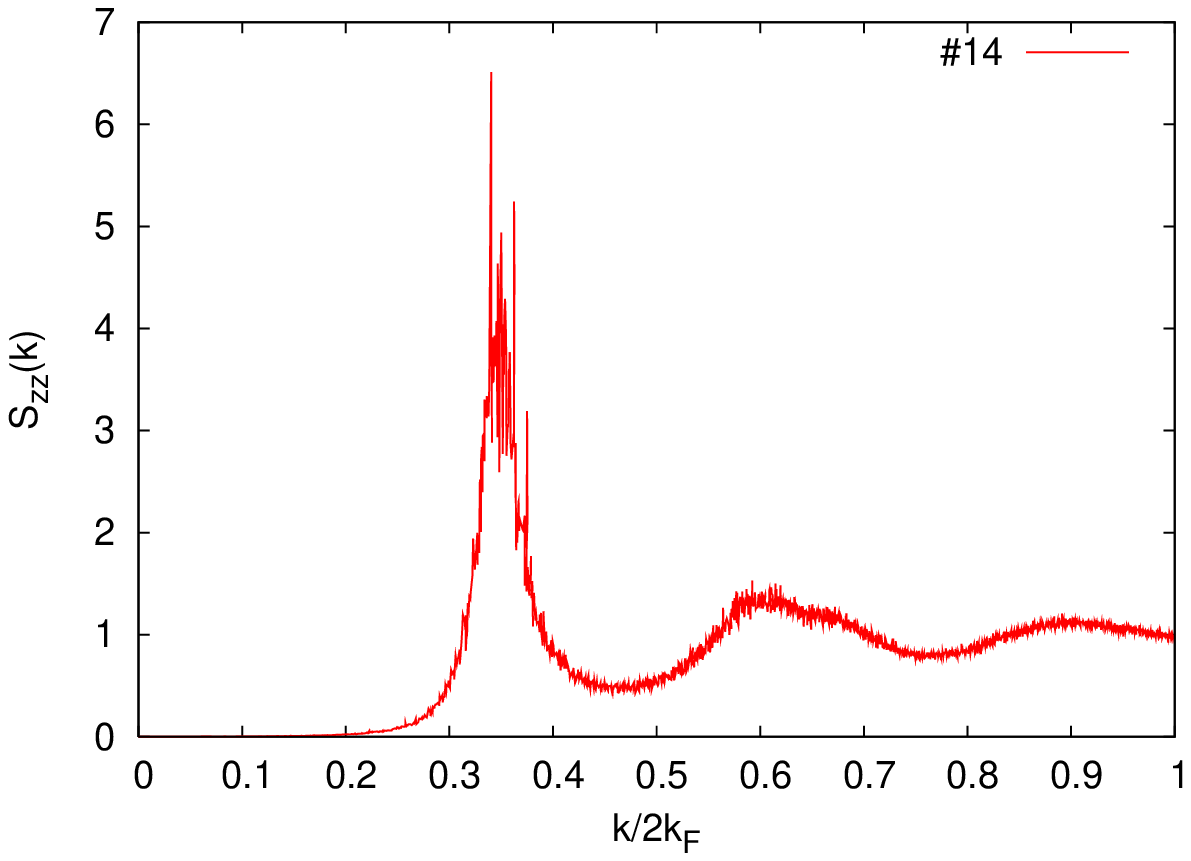}
\includegraphics[scale=.50,angle=0]{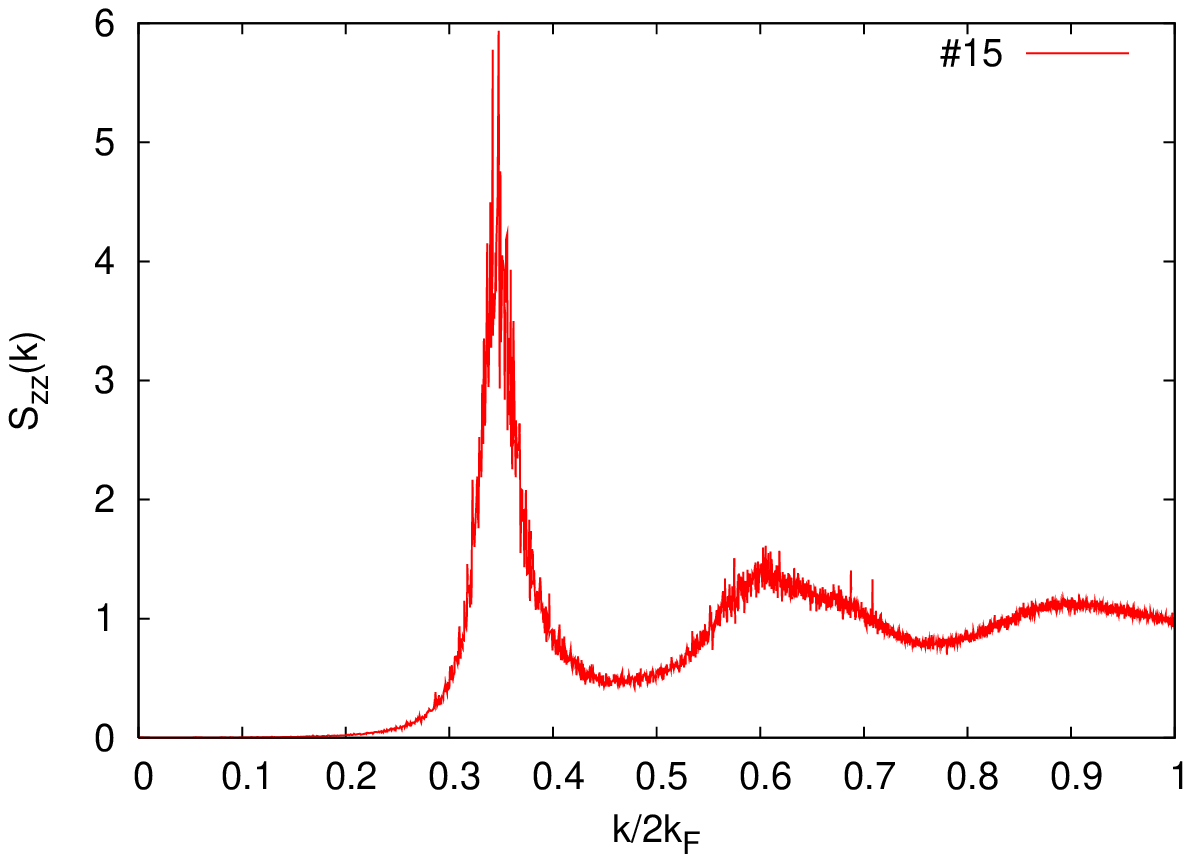}\\
\includegraphics[scale=.50,angle=0]{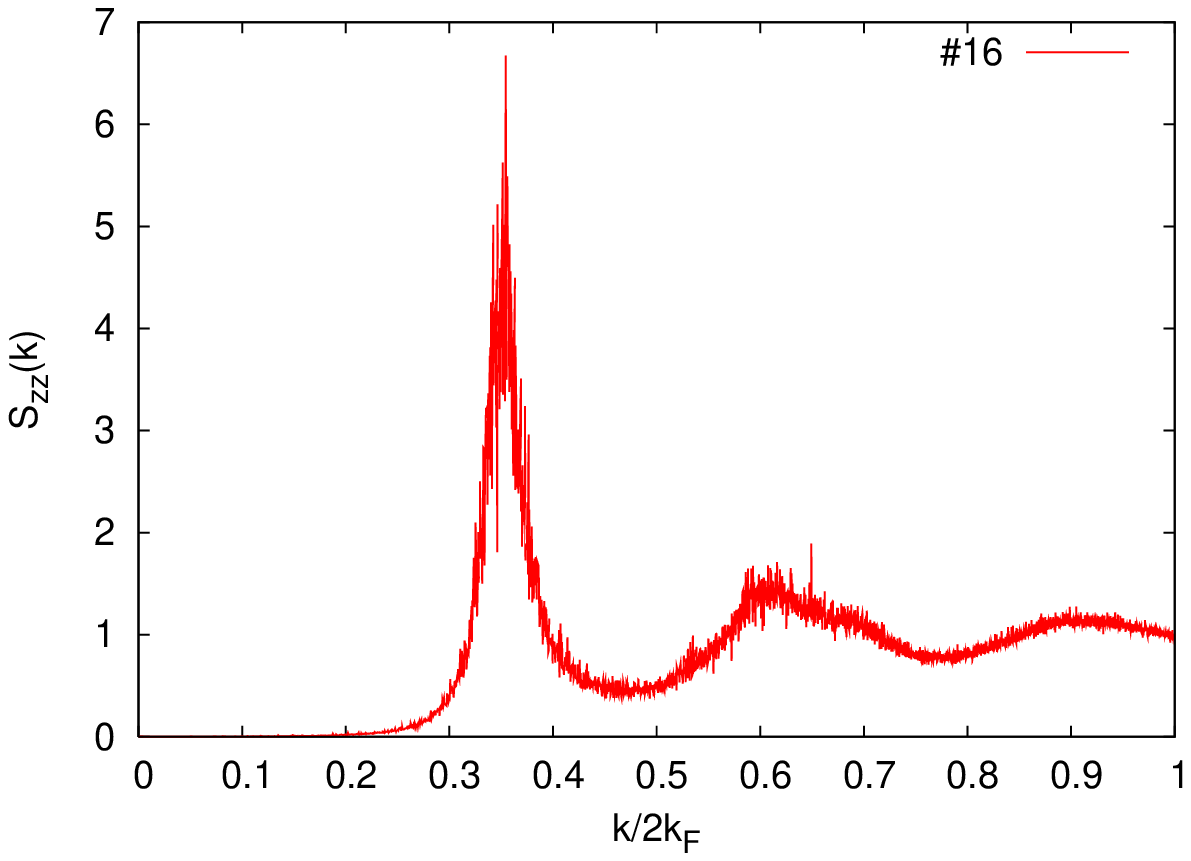}
\includegraphics[scale=.50,angle=0]{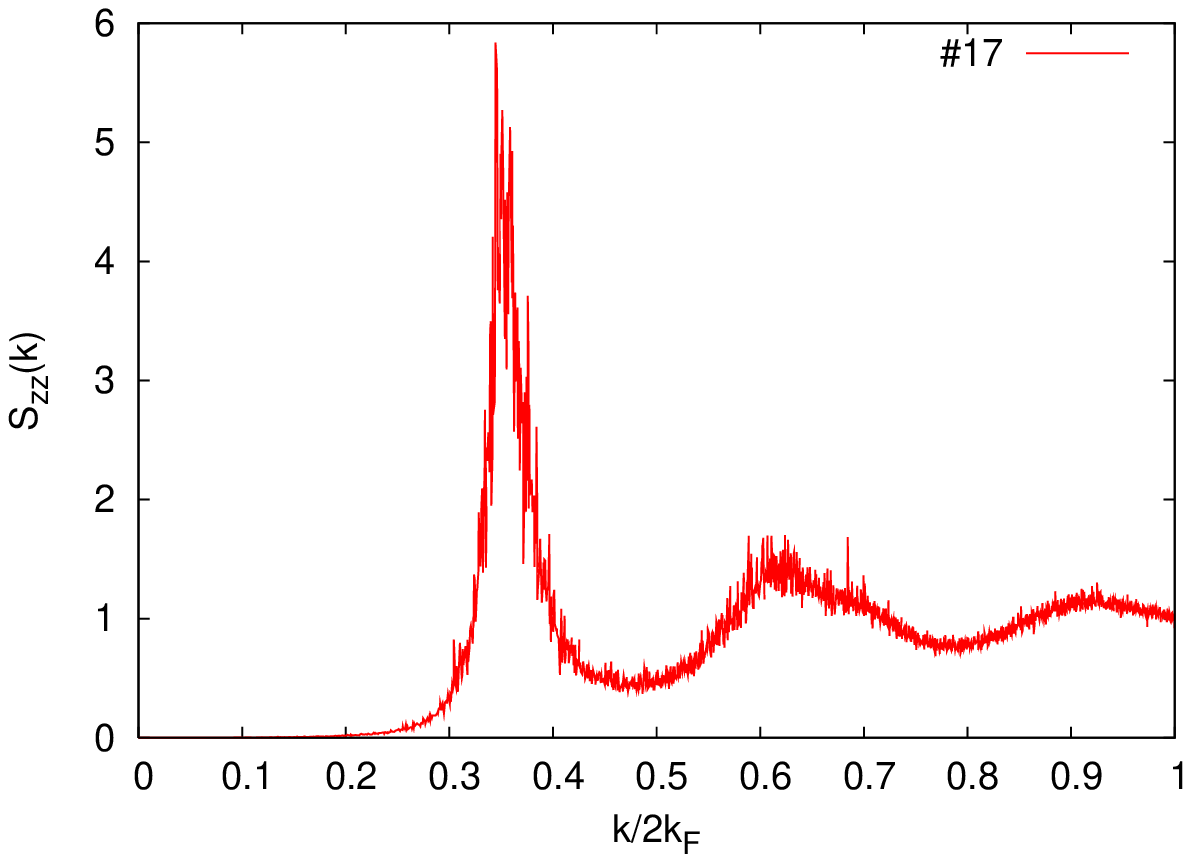}
\caption{(color) The structure factor $S_{\rm zz}(k)$ obtained with MD simulations for the mixtures \#12-17 defined in table \ref{table2}.
\label{figure8}
}
\end{figure}

\begin{figure} 
\includegraphics[scale=.60]{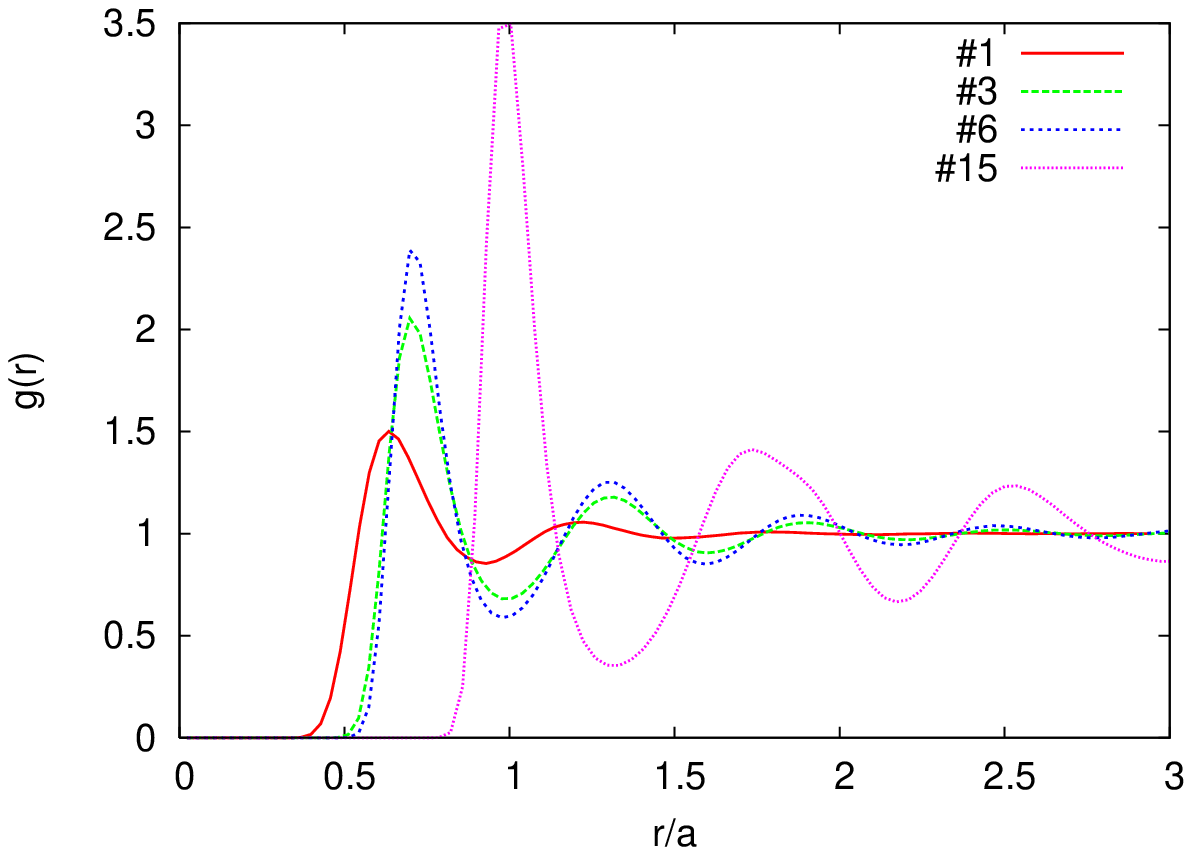}\\
\includegraphics[scale=.60]{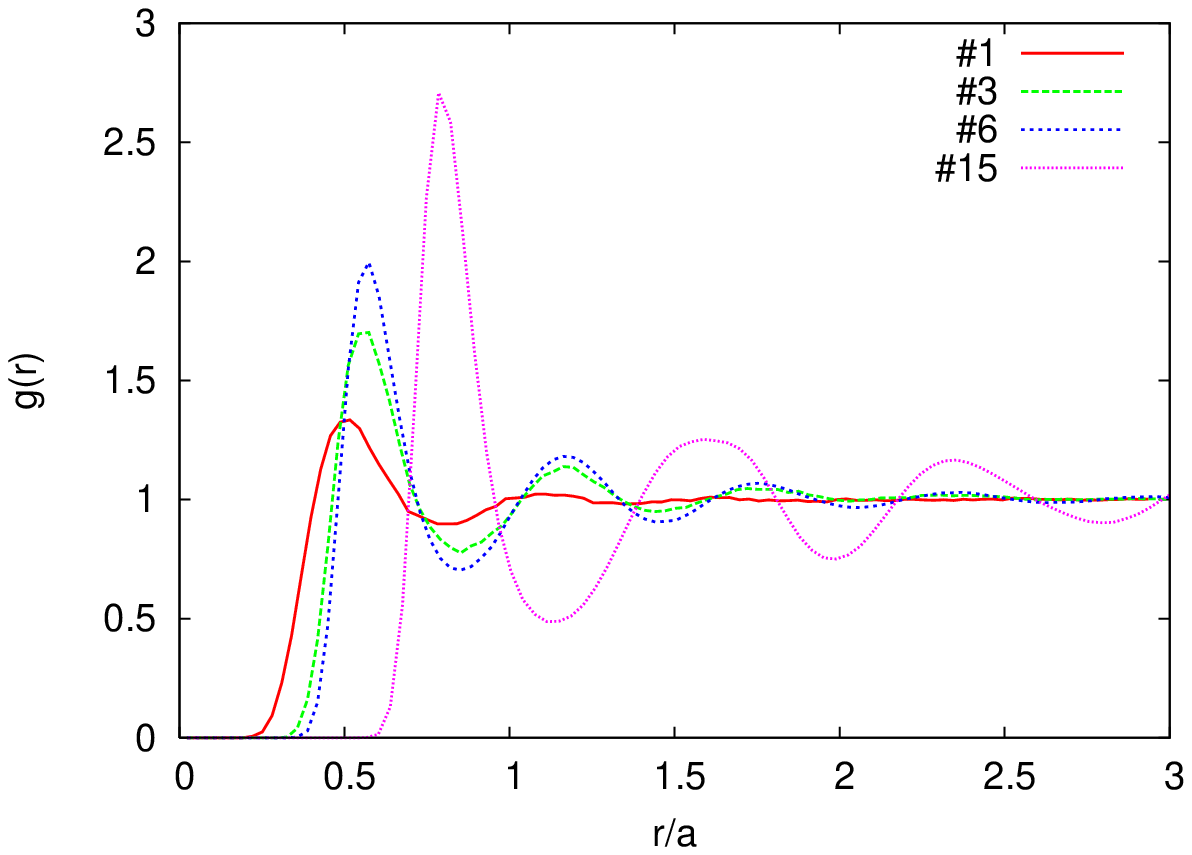}
\caption{(color) (upper panel) Pair distribution function $g(r)$ between particles of most abundant species of compositions \#1, \#3, \#6 and \#15 defined in table \ref{table2}.
(lower panel) Pair distribution function between the most abundant and another less abundant species for the same compositions of the upper panel.
\label{figure14}
}
\end{figure}

The validity of the MLMR (\ref{SofkLMR}) is confirmed by our MD simulations.
The structure factors for the compositions (1) to (8) of table \ref{table2}, corresponding to $8.6 \cdot 10^{6}\leq\rho\leq 4.8 \cdot10^{9}$ $\rm g.cm^{-3}$ and $70\leq\Gamma\leq 140$, are shown in Fig.(\ref{figure1}) and Fig.(\ref{figure2}).
The blue lines show the results obtained using MD simulations and the red lines show the results obtained with the mixing rule Eq.(\ref{Sqqmixing}).
The latter was evaluated with the structure factors for the one-component plasmas obtained by solving the so-called HNC equations (see e.g. \citet{Young})
For all the compositions, Figs.(\ref{figure1}) and (\ref{figure2}) show very good agreement between the MD and the mixing rule values.
In each case, the position of the peaks is well reproduced and the mixing rule overestimates their heights by at most $10\%$.
(The MLMR is also illustrated in the figures of appendix \ref{appendixSk}.)

When used in Eq.(\ref{excess energy}) or Eq.(\ref{CoulombLogarithm}), however, these differences between the MD and MLMR structure factors barely affect the value of the Coulomb logarithm, and in turn the electronic transport properties are unaffected.
This is illustrated in Fig. \ref{figure4} that shows the sum $\ln\Lambda(K)=\int_{0}^{K}{dk\/k^{3}\Big|\frac{v(k)}{\epsilon(k)}\Big|^{2}\left[1-\frac{x_{\rm r}^{2}}{1+x_{\rm r}^{2}}\left(\frac{k}{2k_{\rm F}}\right)^{2}\right]S_{\rm zz}(k)}$ with $0\leq K\leq 2 k_{\rm F}$ obtained with the MD and the mixing rule results.
Thus the MLMR provides a very accurate and convenient way of calculating the electronic transport coefficients for complex ionic mixtures in the liquid phase.
More generally, all quantities that are functionals of the structure factor such as (\ref{excess energy}), the electrical and thermal conductivity and even the NPB emissivities can be obtained to high accuracy from the MLMR.

\subsection{Practical fit to the Coulomb logarithm of a mixture}

Having validated the LMR, we now provide a simple practical fit to the Coulomb logarithm of an arbitrarily complex liquid or amorphous {\it multi-component} plasmas.
The fit is valid for any effective Coulomb coupling $\Gamma_{j}\ge 10$ and relativistic parameter range $1\leq x_{\rm r}\leq 1000$.
For the neutron star crust applications in the amorphous state, once the composition as function of depth is known, the conductivity can be calculated very accurately (when compared to numerical MD simulations) without resort to complicated tabulations of fit coefficients or decomposition of conductivity from different scattering processes.

Our prescription is as follows:
\begin{eqnarray}
\ln\Lambda \approx \frac{\langle Z\rangle^{1/3}}{\langle Z^{2}\rangle}\sum_{j}^{N_{\rm sp}}{x_{j}Z_{j}^{5/3}\ln\Lambda^{\rm OCP}(\Gamma_{j},Z_{j},x_{\rm r})}\/,
\end{eqnarray}
where $\Gamma_{j}=Z_{j}^{2}e^{2}/a_{j}k_{\rm B}T$ and $\ln\Lambda^{\rm OCP}$ is fitted by the expression
\begin{eqnarray}
\lefteqn{\ln\Lambda^{\rm OCP}(\Gamma_{j},Z_{j},x_{\rm r})}&& \label{lambdaprescription}\\
&&\approx C(Z_{\rm j},\Gamma_{\rm j},x_{\rm r})\frac{1+0.177\Gamma_{j}+0.00001\Gamma_{j}^{2}}{1+a(x_{\rm r})\Gamma_{j}+b(x_{\rm r})\Gamma_{j}^{2}}\ln\Lambda_{\rm imp}(x_{\rm r})\/.\nonumber
\end{eqnarray}
In Eq.(\ref{lambdaimp}),
\begin{eqnarray}
\lefteqn{\ln\Lambda_{\rm imp}(x_{\rm r})=\frac{1+396\/x_{\rm r}}{0.466+394\/x_{\rm r}}}&&\label{lnL0}\\
&&\times\frac{1}{2}\left[\left(1+4\beta^{2}q^{2}\right)\ln\left(1+q^{-2}\right)-\beta^{2}-\frac{1+\beta^{2}q^{2}}{1+q^{2}}\right]\/,\nonumber
\end{eqnarray}
is obtained by setting $\epsilon(k)=1+k_{\rm rTF}^{2}/k^{2}$ in Eq.(\ref{lambdaimp}), where
\begin{eqnarray*}
\beta&=&\frac{v_{\rm F}}{c}=\frac{x_{\rm r}}{\sqrt{1+x_{\rm r}^{2}}}\/,\\
q&=&\frac{k_{\rm rTF}}{2k_{\rm F}}=\frac{0.048196}{\sqrt{\beta}}\/,
\end{eqnarray*}
$k_{\rm rTF}$ is the relativistic Thomas-Fermi wavevector, and the functions $a(x_{\rm r})$ and $b(x_{\rm r})$ are given by:
\begin{eqnarray}
a(x_{\rm r})=\left\{
\begin{array}{l}
\frac{1+a_{1}\/x_{\rm r}^{1/3}}{a_{2}+a_{3}\/x_{\rm r}^{3/2}}\quad x_{\rm r}<5 \label{fit_a}\\
\\
2.22  \quad x_{\rm r}\ge 5
\end{array}
\right.
\end{eqnarray}
\begin{eqnarray}
b(x_{\rm r})=\left\{
\begin{array}{l}
\frac{1+b_{1}/x_{\rm r}^{2}}{b_{2}+b_{3}/x_{\rm r}^{2.3}}\quad x_{\rm r}<10 \label{fit_b}\\
\\
0.087  \quad x_{\rm r}\ge 10
\end{array}
\right.
\end{eqnarray}
with the parameters $a_{n}$ and $b_{n}$ as listed in table \ref{table5}.

The species dependence is encapsulated in the function
\begin{eqnarray}
C(Z_{\rm j},\Gamma_{\rm j},x_{\rm r})=\left\{
\begin{array}{l}
1\,,\quad Z_{\rm j}=1\\
\\
C_{0}(Z_{\rm j},x_{\rm r})+C_{1}(Z_{\rm j},x_{\rm r})\Gamma_{j} \quad Z_{\rm j}>1\/,
\end{array}
\right.
\end{eqnarray}
where
\begin{eqnarray*}
C_{0}(Z_{\rm j},x_{\rm r})&=&a_{0}(x_{\rm r})+a_{1}(x_{\rm r})\ln\/Z_{j}\\
C_{1}(Z_{\rm j},x_{\rm r})&=&b_{0}(x_{\rm r})+b_{1}(x_{\rm r})\ln\/Z_{j}\/.
\end{eqnarray*}
We fit the coefficients $a_{0}(x_{\rm r}), a_{1}(x_{\rm r}), b_{0}(x_{\rm r}), b_{1}(x_{\rm r})$ piecewise over disjoint $\Gamma_{j}$ and $x_{r}$ ranges as follows:
For $10\leq\Gamma_{j}\leq 30$:
\begin{eqnarray*}
a_{0}(x_{\rm r})&=&\left\{
\begin{array}{l}
0.973587-0.0790899\/\ln\/x_{\rm r}\quad x_{\rm r}<10\\
\\
0.806877  \quad x_{\rm r}\ge 10
\end{array}
\right.\\
a_{1}(x_{\rm r})&=&\left\{
\begin{array}{l}
0.652748+0.117973\/\ln\/x_{\rm r}\quad x_{\rm r}<2.5\\
\\
0.766772  \quad x_{\rm r}\ge 2.5
\end{array}
\right.\\
b_{0}(x_{\rm r})&=&-0.029204+0.0367777\exp(-0.93315\/\ln\/x_{\rm r})\\
b_{1}(x_{\rm r})&=&0.0853907-0.053054\exp(-0.885854\/\ln\/x_{\rm r})\/,
\end{eqnarray*}
whereas for $30<\Gamma_{j}$:
\begin{eqnarray*}
a_{0}(x_{\rm r})&=&0.979517-0.1279\/\ln\/x_{r}\\
&&+0.0252858\/(\ln\/x_{r})^{2}-0.00204823\/(\ln\/x_{r})^{3}\\
a_{1}(x_{\rm r})&=&\left\{
\begin{array}{l}
1.02454+0.0577405\/\ln\/x_{\rm r}\quad x_{\rm r}<5\\
\\
1.09301+0.0135052\/\ln\/x_{\rm r}\quad x_{\rm r}\ge 5
\end{array}
\right.\\
b_{0}(x_{\rm r})&=&\left\{
\begin{array}{l}
0.00841864-0.0124097\/\ln\/x_{\rm r}\quad x_{\rm r}<10\\
\\
-0.0193203  \quad x_{\rm r}\ge 10
\end{array}
\right.\\
b_{1}(x_{\rm r})&=&\left\{
\begin{array}{l}
0.0225627+0.0242463\/\ln\/x_{\rm r}\quad x_{\rm r}<5\\
\\
0.0422014+0.0117571\/\ln\/x_{\rm r}  \quad 5<x_{\rm r}<10\\
\\
-0.0193203  \quad 10<x_{\rm r}
\end{array}
\right.
\end{eqnarray*}

This fit was obtained to match the OCP structure factors of \citet{Young} for the range  $5\leq\Gamma_{j}<225$.
However, assuming that the system stays amorphous (no periodic lattice structure), the fit has not restriction on $\Gamma_{j}$ as demonstrated in Fig.\ref{figure17}.

\section{Application of the  MLMR to obtain thermal and electrical conductivities of superburst progenitors } \label{section3}

Figure (\ref{figure16}) shows the thermal conductivity calculated from numerical MD simulations in this paper and accurately reproduced by using the mixing rule prescription of Eq.(\ref{prescription2009}).
The conductivity of the amorphous outer crust is lower by an order of magnitude from estimates obtained using the impurity parameter formalism, Eq.(\ref{nuimp}).
Figure (\ref{figure16}) (see also Fig.(\ref{figure1})) also show that $\langle Z^{2}\rangle$, and as a consequence the conductivity, are less sensitive than $Q_{\rm imp}$ to reaction processes that reduce crust impurity abruptly in the outer crust, such as (1) electron captures on very proton-rich nuclei produced in the rp-process (and therefore far away from beta-stability at the top of the crust), (2) charge particle capture and fusion reactions destroying lighter nuclei, and (3) $(\gamma,n)$ and $(n,\gamma)$-driven rearrangement of abundances in mass chains. 

We note that in the limit of a classical amorphous solid, the Coulomb logarithms for electrical and thermal conductivity are equal, i.e.
$\ln  \Lambda_\sigma = \ln \Lambda_\kappa$, therefore $\sigma$ and $\kappa$ are related by the Wiedemann-Franz law, Eq.(\ref{WF}).
Thus while we only show a plot of the outer crust thermal conductivity as function of density in fig.(\ref{figure16}), the composition-dependent electrical conductivity is related very
simply to it.

X-ray burst ash composition differences will exist between 1-zone models \citep{SchatzXRB} of an X-ray Burst as opposed to a multi-zone model \citep{Woosley2004}. Our purpose here is not to study all possible compositions in a superburst progenitor, but rather to show the applicability of the Microscopic Linear Mixing Rule to an arbitrarily complex composition with the heterogeneity expected of X-ray burst ashes. Once composition is known, MD permits us to determine the physical state and the conductivity of crust matter. Since the thermal conductivity is a critical ingredient setting the crust thermal gradient, an uncertainty in this quantity makes it difficult to predict whether shallow crust conditions $\sim10^9$ g/cc are conducive to
superburst ignition. We will leave the full thermal profile calculation to a later work, but we point out here that a conductivity significantly lowered by impurities, such as the MD results show for the outer crust which is in the amourphous outer crust can foster a steep thermal gradient such as would be required for superburst ignition.
In \citet{Brown2004} a local stability analysis for accretion rate $\dot{M}=3 \cdot 10^{17}$ g s$^{-1}$  shows that superburst ignition at column depth $y \approx 10^{12}$ g cm${^-2}$ is favored by an amorphous crust. Further, such a crust renders the thermal profile relatively insensitive to the core neutrino luminosity, whether from modified URCA processes (inefficient core cooling) or enhanced due to direct URCA or pionic reactions (efficient core cooling). These results were confirmed by the global stability analysis of \citet{CooperNarayan2005}.

A complete analysis of thermal structure requires an understanding of conductivity in the inner crust and therefore of nuclear processes at neutron drip which may reduce impurity significantly (but not completely) from that of X-ray burst ashes. We are currently performing these detailed studies and will publish the results shortly.

\acknowledgments
We would like to thank Sanjay Reddy and Dany Page for stimulating discussions.
This work was performed for the U.S. Department of Energy by Los Alamos National Laboratory under contract DE-AC52-06NA25396.



\appendix

\section{Details of MD simulations} \label{appendix1}

Some technical details of the MD simulations are collected in table \ref{table3}, namely
\begin{itemize}
\item $N_{\rm sp}=$ number of nuclear species;
\item $N=$ total number of particles in the MD simulations;
\item $N_{\rm eq}=$ number of time steps used initially to let the system equilibrate;
\item $N_{\rm run}=$ number time steps of the MD run after equilibration;
\item $N_{\rm k}=$ total number of wave-vector norms used to calculated $S_{\rm zz}(k)$ with $a_{1}k\leq 22$, see Eqs.(\ref{kMD}-\ref{SzzMD}).
\end{itemize}
In all the simulations, the time step $\delta t$ is chosen such that $\delta t=0.01/\omega_{\rm p}$, where $\omega_{\rm p}$ is the mean ion plasma frequency.
The total energy is very well conserved during the simulation (better than one part in $10^{7}$).

\section{Relativistic dielectric function}\label{appendix2}

The Random-Phase Approximation static dielectric function of the relativistic electron gas in its ground state is \citep{Jancovici1962}
\begin{eqnarray}
\epsilon(k)&=&1+\frac{k_{\rm TF}^{2}}{k^{2}}\left\{\frac{2}{3}\sqrt{1+x_{\rm r}^{2}}-\frac{2x_{\rm r}}{3}x^{2}{\rm sinh}^{-1}x_{\rm r}+\sqrt{1+x_{\rm r}^{2}}\frac{1+x_{\rm r}^{2}-3x_{\rm r}^{2}x^{2}}{6x_{\rm r}^{2}\/x}\ln\Big|\frac{1+x}{1-x}\Big|\right.\\
&&\hspace{2cm}\left.-\frac{1-2x_{\rm r}^{2}x^{2}}{6x_{\rm r}^{2}x}\sqrt{1+x_{\rm r}^{2}x^{2}}\ln\Big|\frac{\sqrt{1+x_{\rm r}^{2}x^{2}}+x\sqrt{1+x_{\rm r}^{2}}}{\sqrt{1+x_{\rm r}^{2}x^{2}}-x\sqrt{1+x_{\rm r}^{2}}}\Big|\right\}\/, \label{epsilon_e}
\end{eqnarray}
where $x=k/2k_{\rm F}$ and $k_{\rm TF}$ is the (nonrelativistic) Thomas-Fermi (TF) inverse screening length
\begin{eqnarray}
k_{\rm TF}=\frac{e\sqrt{12\pi\/m_{e}\/n_{e}}}{\hbar k_{\rm F}} \label{kTF}
\end{eqnarray}

\section{Additional structure factors} \label{appendixSk}

Fig.(\ref{figure12}) and (\ref{figure13}) shows the charge-charge structure factors $S_{\rm zz}(k)$ and their mixing rule approximation for some of the compositions of table \ref{table2} but temperatures different than $0.5$ GK, as listed in table \ref{table3}.

\clearpage

\clearpage

\clearpage

\clearpage

\clearpage

\clearpage

\clearpage

\clearpage

\clearpage

\begin{figure} 
\includegraphics[scale=.50]{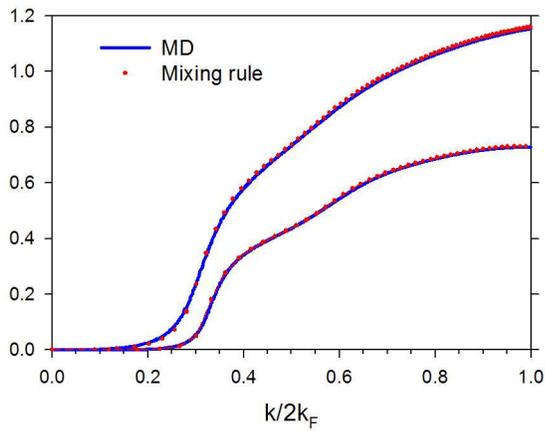}
\caption{(color) Cumulated integral $\ln\Lambda(K)=\int_{0}^{K}{dk\/k^{3}\Big|\frac{v(k)}{\epsilon(k)}\Big|^{2}\left[1-\frac{x_{\rm r}^{2}}{1+x_{\rm r}^{2}}\left(\frac{k}{2k_{\rm F}}\right)^{2}\right]S_{\rm zz}(k)}$ obtained using the $S_{\rm zz}(k)$ from MD simulations (full blues line) and from the mixing rule Eq.(\ref{Sqqmixing}) (red dots). The Coulomb logarithm Eq.(\ref{CoulombLogarithm}) is $\ln\Lambda(2k_{\rm F})$. The upper and lower curves correspond to the mixture (1) and (9) defined in table (\ref{table2}), respectively.
\label{figure9}
}
\end{figure}

\clearpage

\begin{figure} 
\includegraphics[scale=.60]{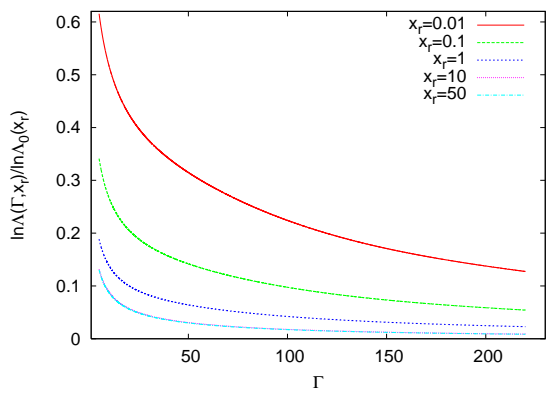}
\caption{OCP Coulomb logarithm as a function of the Coulomb coupling $\Gamma$ for several values of the relativistic parameter $x_{\rm r}$.
\label{figure15}
} 
\end{figure}

\clearpage

\begin{figure} 
\includegraphics[scale=1]{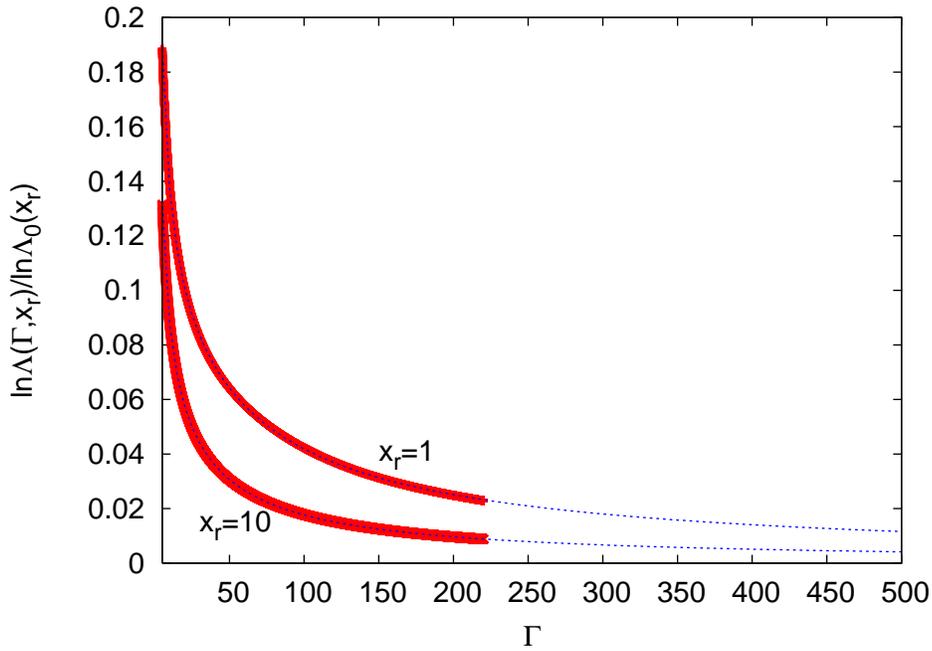}
\caption{OCP Coulomb logarithm $\ln\Lambda^{\rm OCP}$ at $x_{\rm r}=1$ and $x_{\rm r}=10$.
The red solid lines are obtained from Eq.(\ref{OCPCOulomblogarithm}) with the OCP structure factors of \citet{Young} valid for $5\leq \Gamma\leq 225$.
The blue dotted lines are from our prescription Eq.(\ref{lambdaprescription}) and fit the values obtained with \citet{Young} very well and smoothly extend into the $\Gamma>225$ regime.
\label{figure17}
}
\end{figure}

\clearpage

\begin{figure} 
\includegraphics[scale=.60]{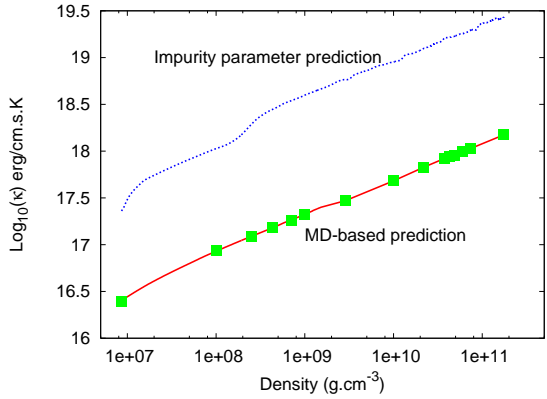}
\caption{Thermal conductivity as a function of density in the crust of an accreting neutron star with $\dot{M}\approx0.1\dot{M}_{\rm Edd}$. The molecular dynamics (MD) calculations (squares) which are accurately reproduced by the linear mixing rule (\ref{prescription2009}) (red line) are contrasted with the electron-impurity scattering conductivity as calculated using Eq.(\ref{nuimp}) over a range of four orders of magnitude in crust density.
As discussed in section \ref{section3}, $\langle Z^{2}\rangle$, and as a consequence the conductivity, is less sensitive than $Q_{\rm imp}$ to composition changes at densities $\sim 10^{7}$, $2\times 10^{8}$, $10^{10}$ and $10^{11}$ $\rm g.cm^{-3}$.
\label{figure16}
} 
\end{figure}

\clearpage

\begin{figure} 
\includegraphics[scale=.22]{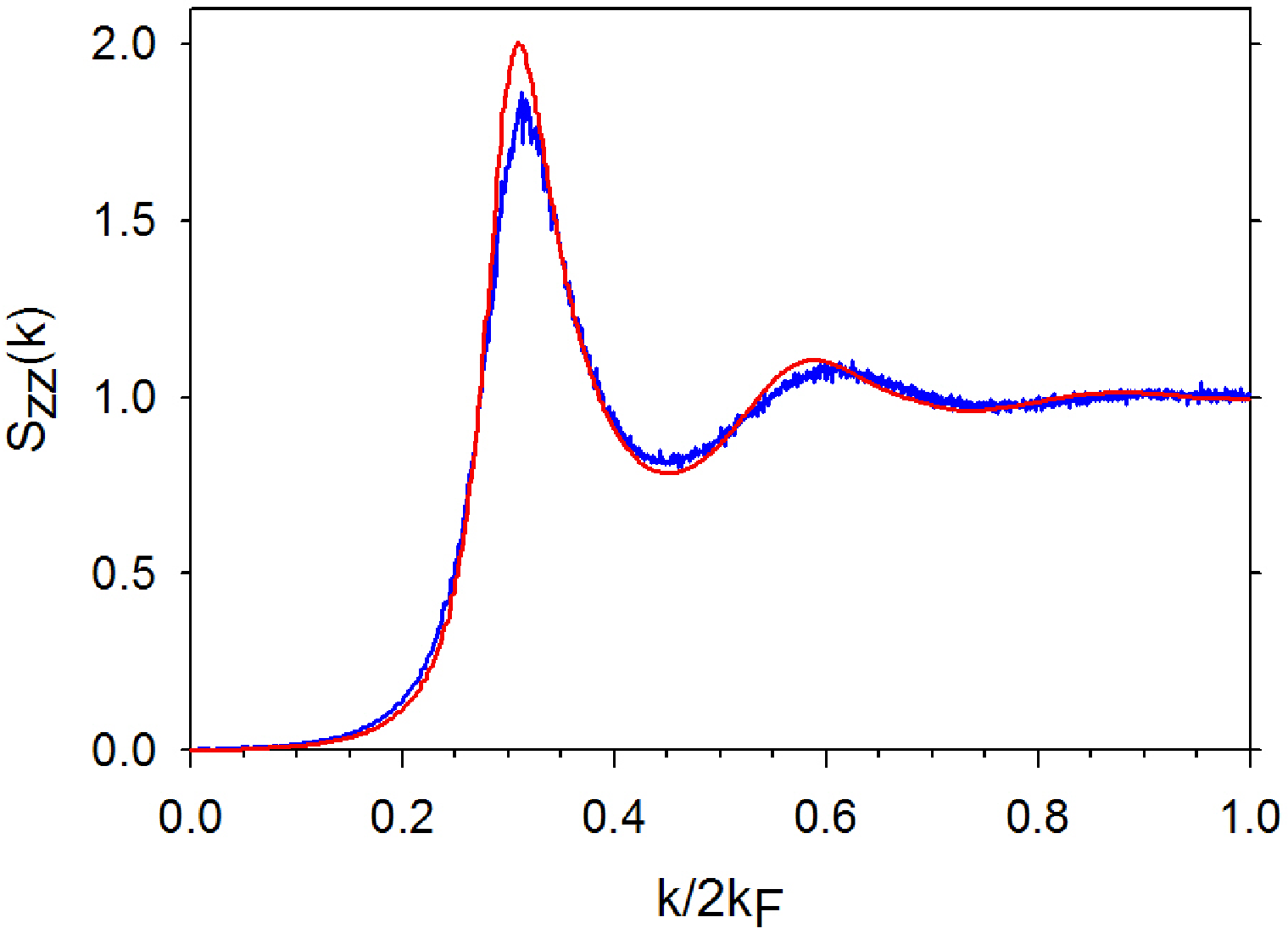}
\includegraphics[scale=.22]{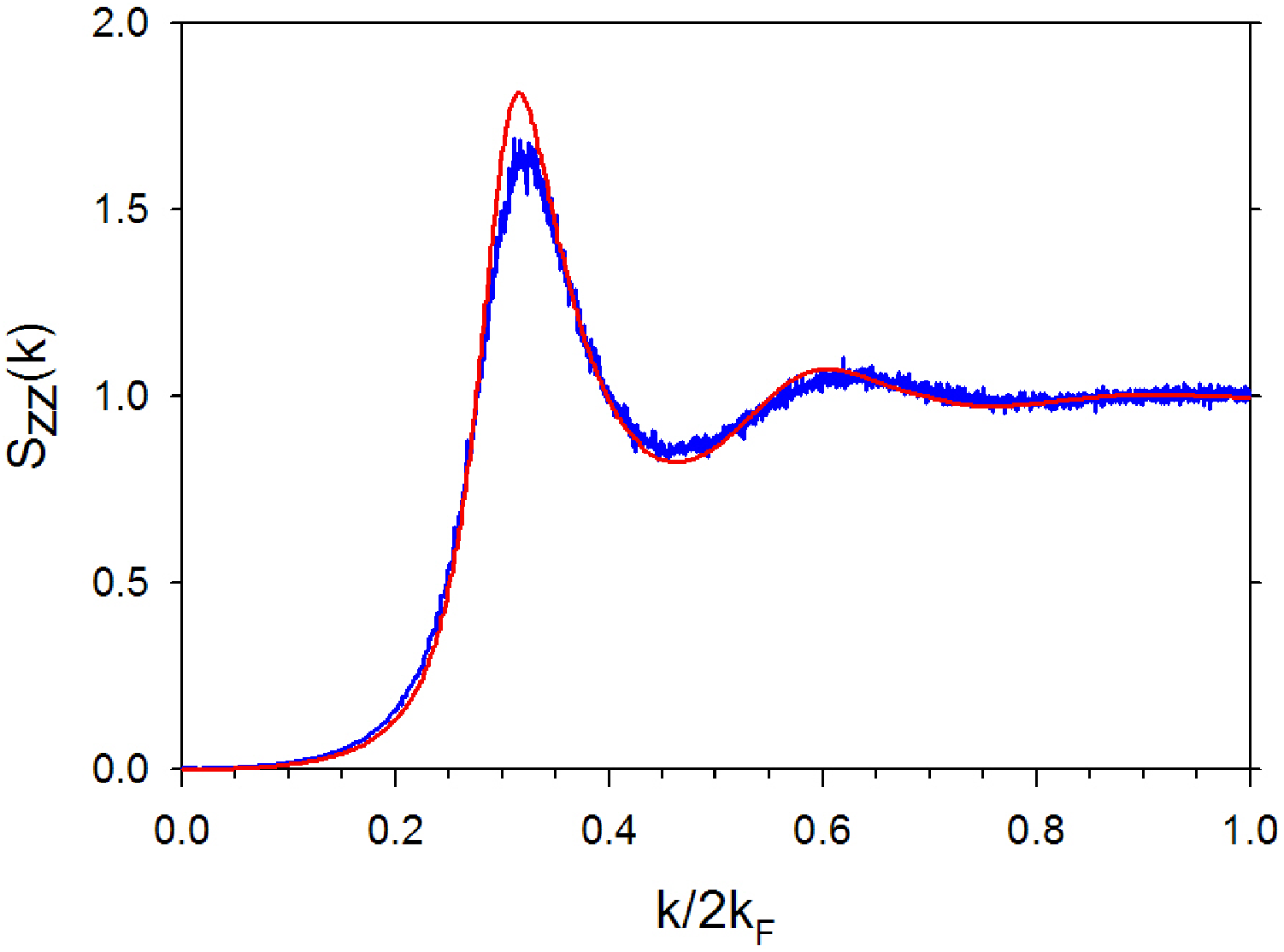}\\
\includegraphics[scale=.22]{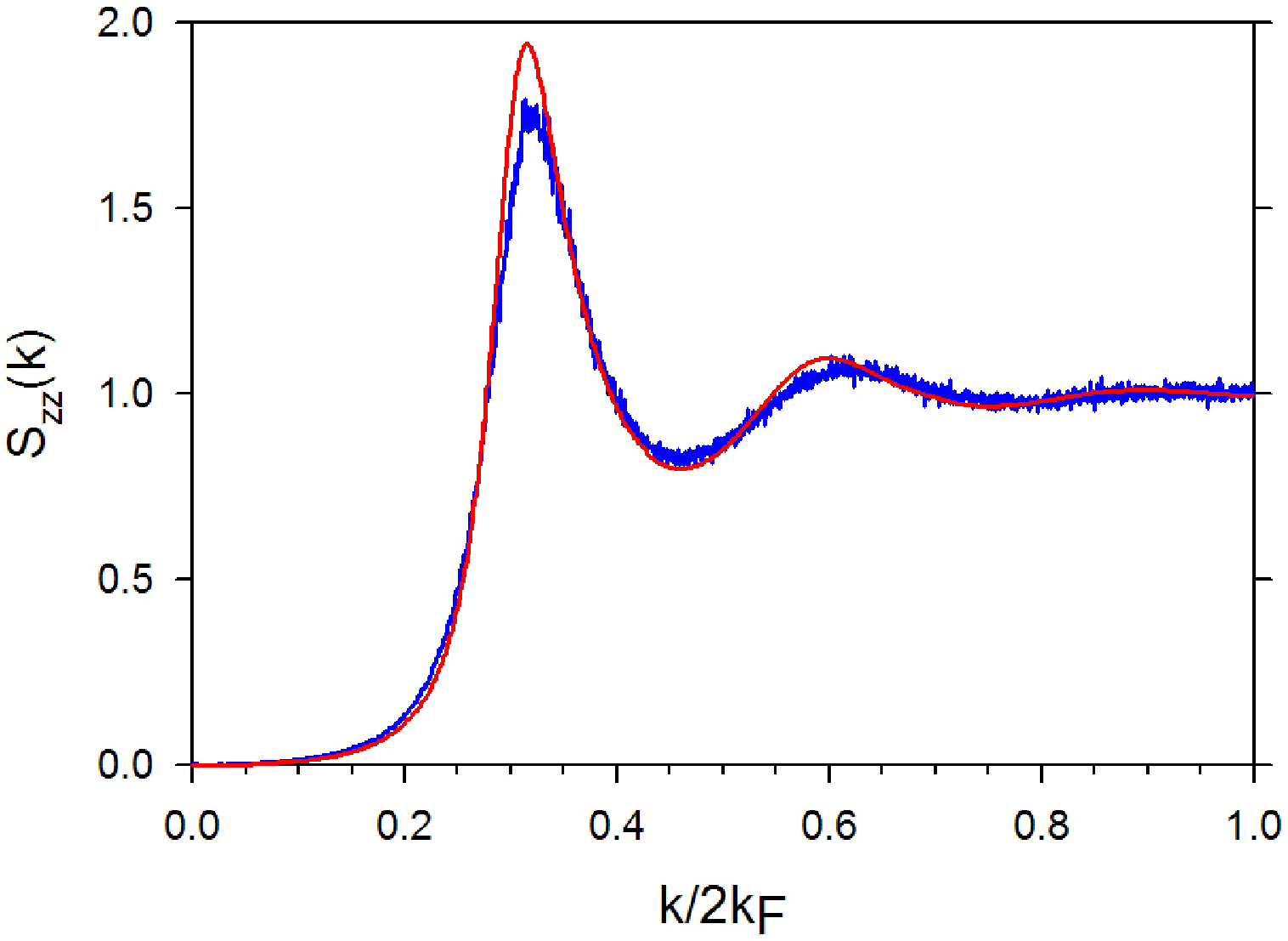}
\includegraphics[scale=.22]{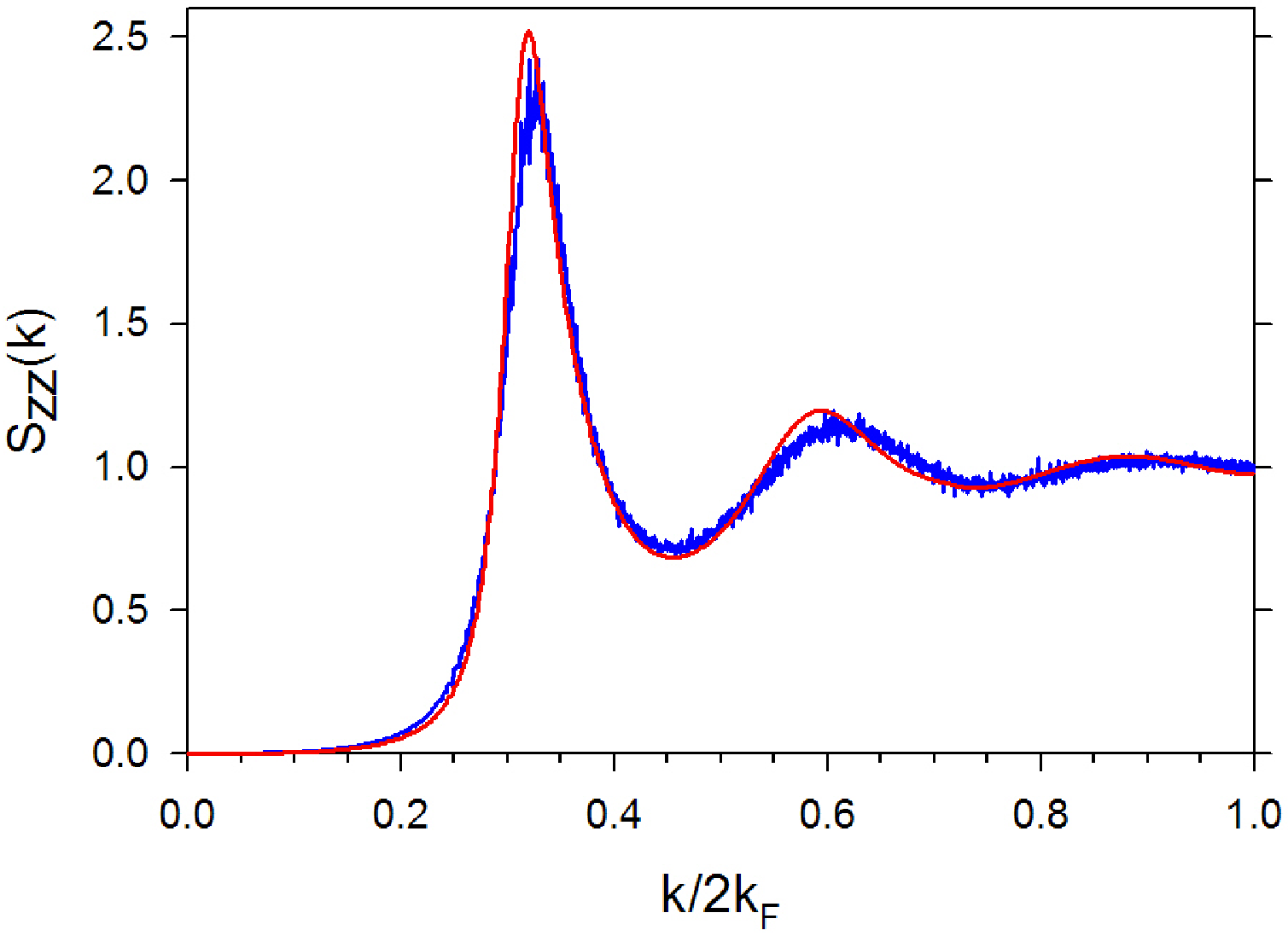}
\caption{(color) The structure factor $S_{\rm zz}(k)$ obtained with MD simulations (blue lines) and
with the mixing rule formula (\ref{SofkLMR}) (red line) for the mixtures \#1-4 defined in table \ref{table4}.
\label{figure12}
}
\end{figure}

\begin{figure}
\includegraphics[scale=.22]{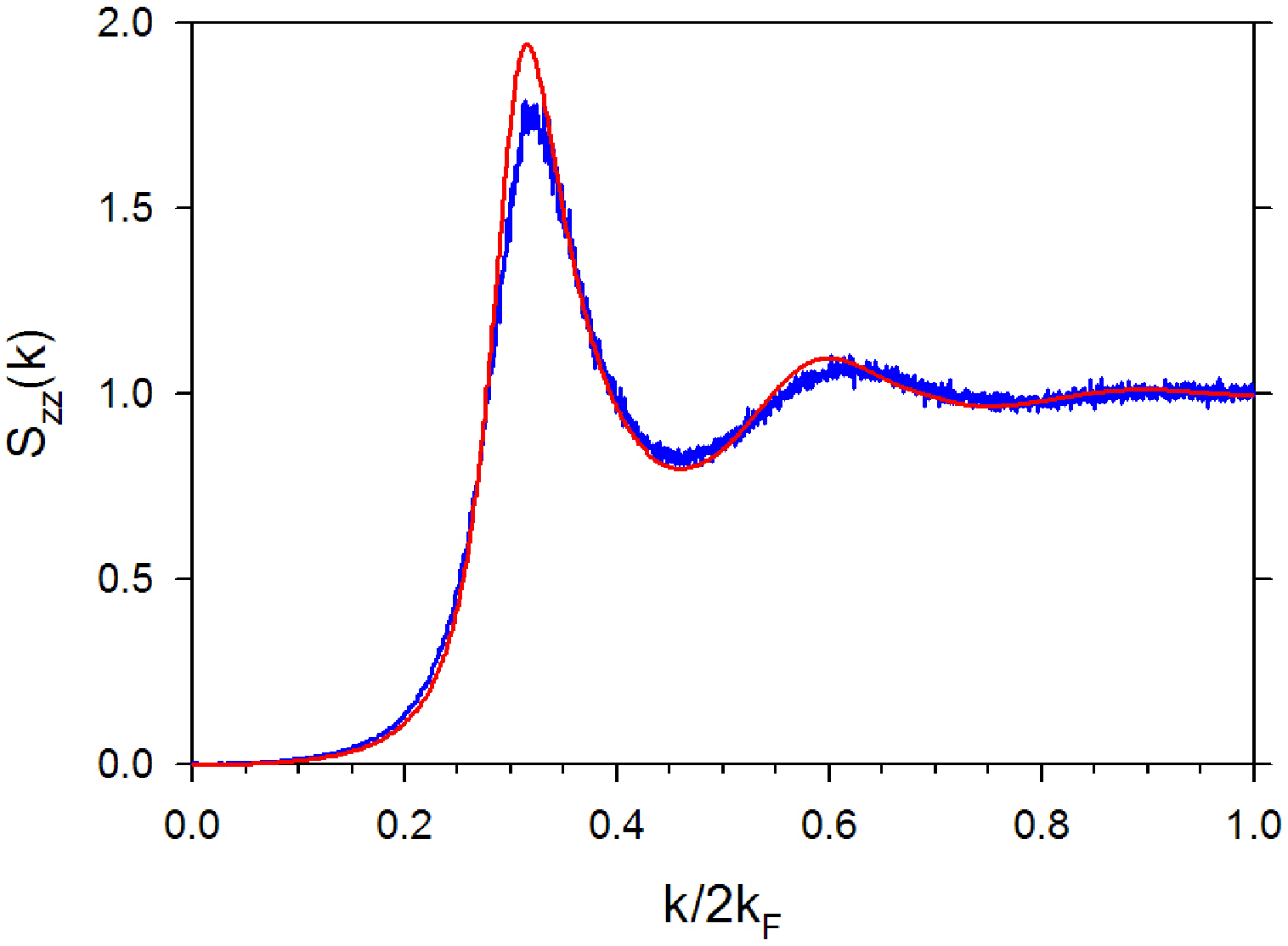}
\includegraphics[scale=.22]{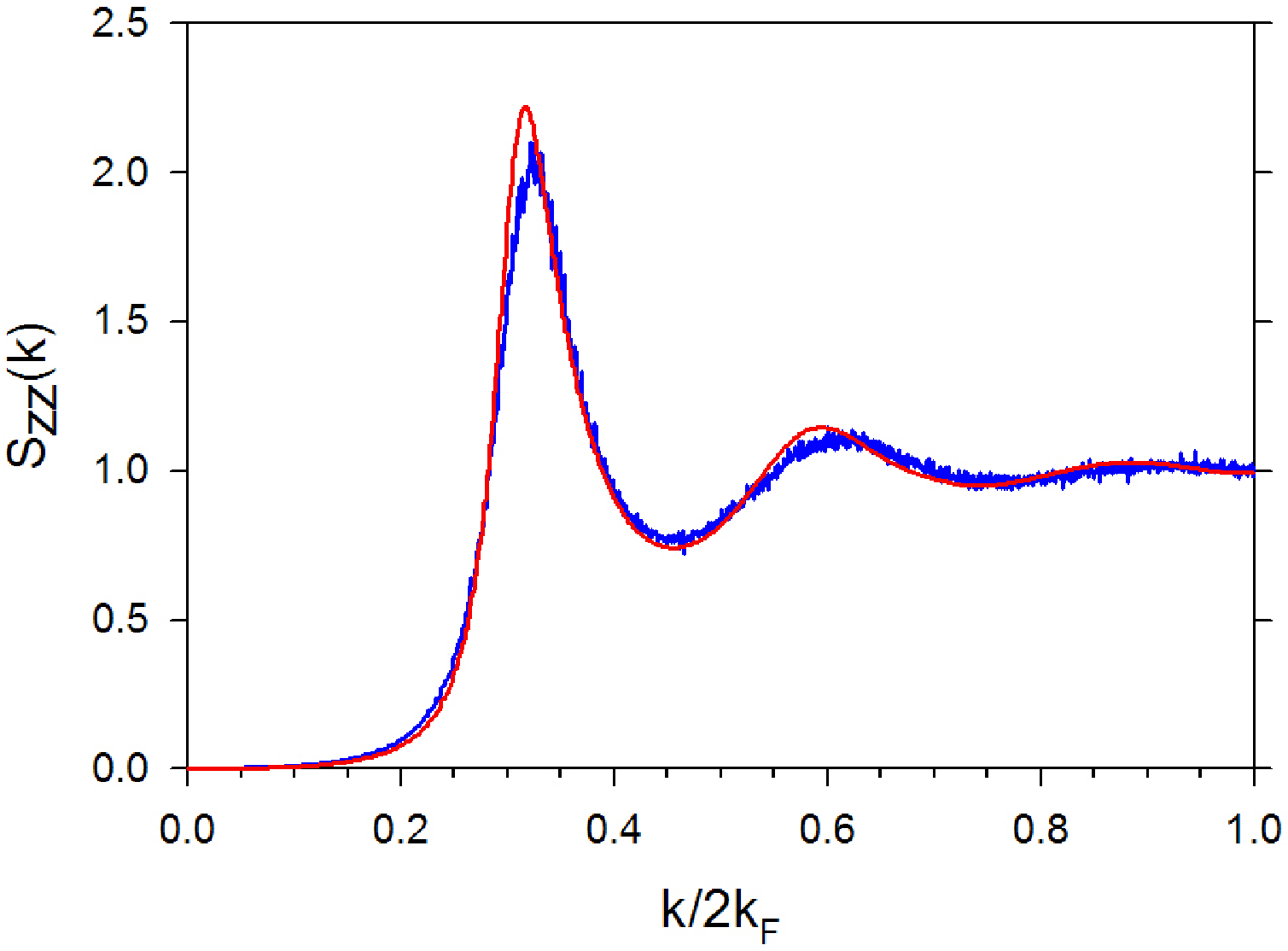}\\
\includegraphics[scale=.22]{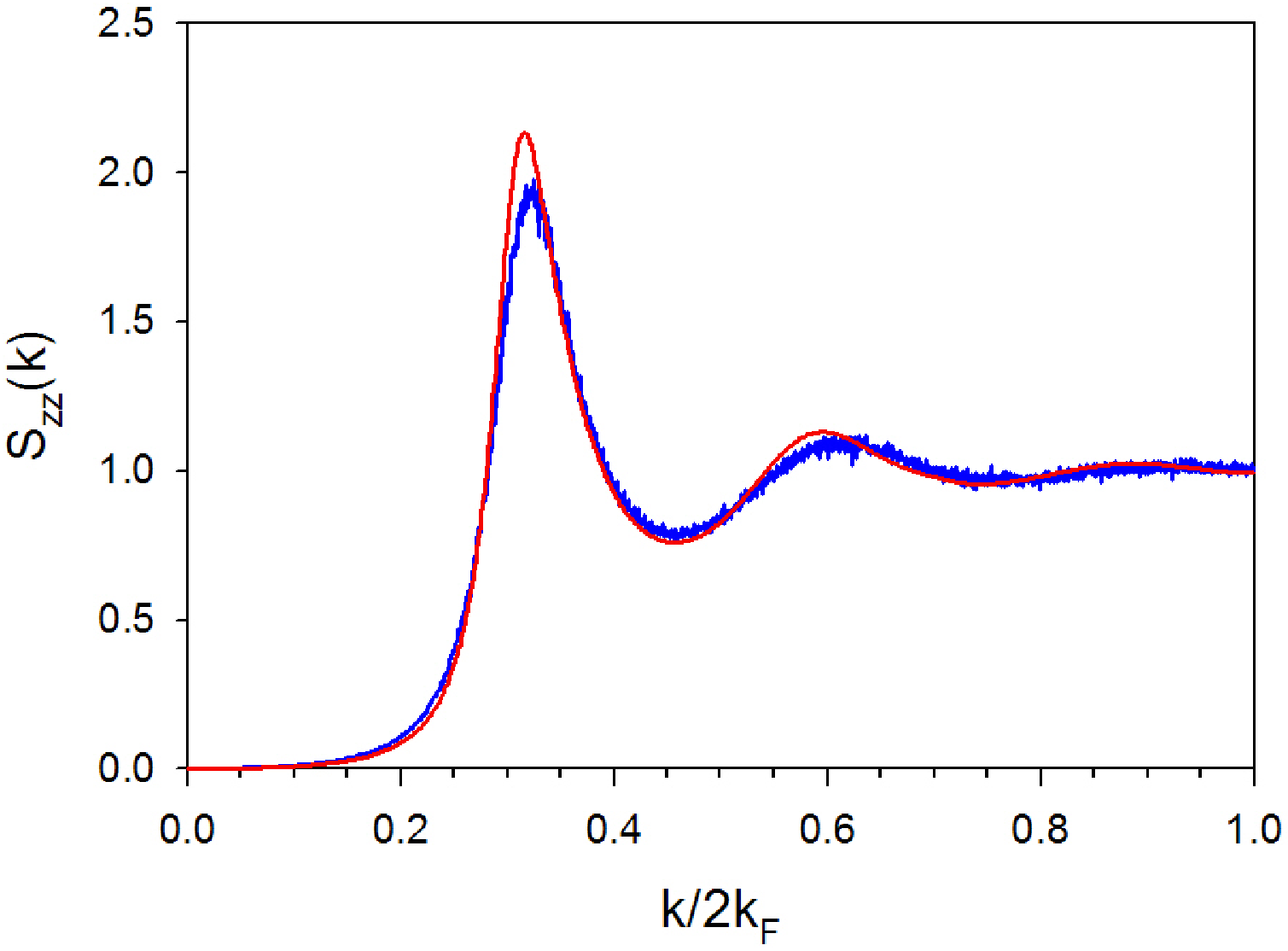}
\includegraphics[scale=.22]{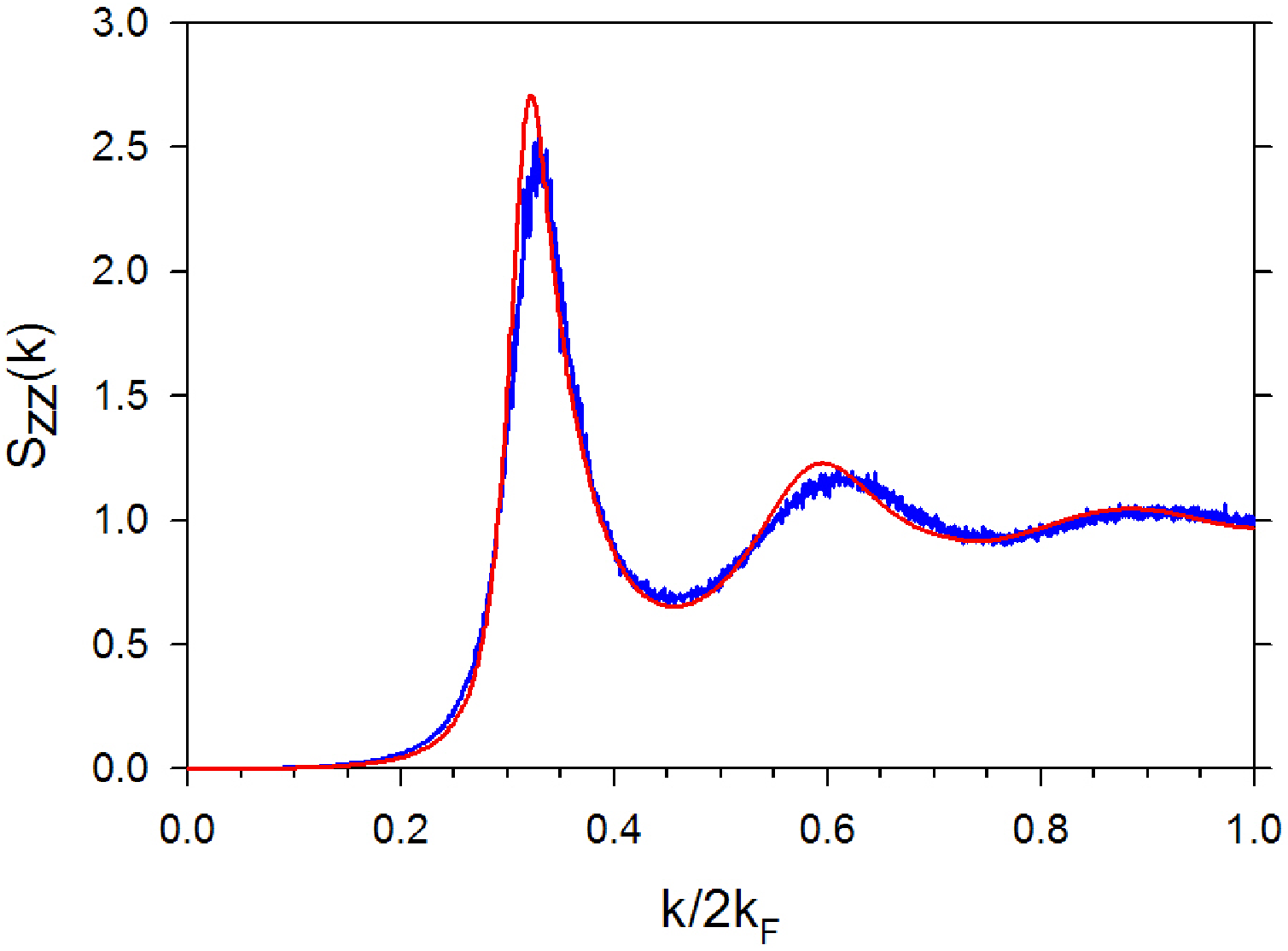}
\caption{(color) Same as figure \ref{figure12} for composition  \#5-8 listed in table \ref{table4}
\label{figure13}
}
\end{figure}

\clearpage

\begin{deluxetable}{ccc}
\rotate
\tablecaption{Entropy variation $\Delta s=\Delta S/nV\/k_{B}\/T$ on melting (i.e. from BCC to fluid) of the OCP as a function of the inverse dimensionless screening length $\kappa=ak_{\rm sc}$.
\label{table1}}
\tablewidth{0pt}
\tablehead{
\colhead{$\kappa$} & \colhead{$\Gamma_{m}$} & \colhead{$\Delta s$}
}
\startdata
0 & 171.8 & 0.8516\\
0.2 & 173.5 & 0.8526\\
0.4 & 178.6 & 0.8551\\
0.6 & 187.1 & 0.8580\\
0.8 & 199.6 & 0.8599\\
1. & 217.4 & 0.8591
\enddata
\tablecomments{$\Gamma_{m}$ is the coupling paramater $\Gamma_{\rm OCP}$ at melting. The data were calculated using the equation of state of \citet{Hamaguchi1997}.}
\end{deluxetable}

\clearpage

\begin{deluxetable}{ccccccccccc}
\rotate
\tablecaption{List of the mixtures considered in this work extracted from \citet{Gupta2007}.
\label{table2}}
\tablewidth{0pt}
\tablehead{
\colhead{mixture} & $N_{\rm sp}^{\rm orig}$ & \colhead{$N_{\rm sp}$} & \colhead{$\rho$ ($\rm g.cm^{-3}$)} & \colhead{$E_{\rm F}$ (MeV)} & \colhead{$ak_{\rm F}$} & \colhead{$\Gamma_{\rm eff}=\langle Z^{5/3}\rangle\Gamma_{e}$} & \colhead{$\langle Z\rangle$} & \colhead{$\langle Z^{2}\rangle$}
}
\startdata
1 & 450 & 55 & 8.59 $10^{6}$ & 0.45 & 6.44 & 32.3 & 37 & 1597\\
2 & 419 & 50 & 1.01 $10^{8}$ & 1.38 & 6.46 & 71.3 & 37 & 1537\\
3 & 451 & 52 & 2.49 $10^{8}$ & 2.00 & 6.44 & 97.9 & 38 & 1564\\ 
4 & 432 & 52 & 4.33 $10^{8}$ & 2.49 & 6.52 & 118.4 & 38 & 1572\\ 
5 & 405 & 50 & 7.08 $10^{8}$ & 3.00 & 6.42 & 137.4 & 38 & 1549\\ 
6 & 400 & 53 & 1.00 $10^{9}$ & 3.42 & 6.41 & 153.2 & 38 & 1536\\ 
7 & 398 & 53 & 2.87 $10^{9}$ & 5.00 & 6.37 & 179.5 & 35 & 1359\\ 
8 & 406 & 52 & 4.85 $10^{9}$ & 6.01 & 6.31 & 236.4 & 36 & 1401\\ 
9 & 411 & 49 & 1.00 $10^{10}$ & 7.72 & 6.26 & 263.1 & 34 & 1275\\
10 & 404 & 49 & 2.17 $10^{10}$ & 10.0 & 6.20 & 343.8 & 34 & 1171\\ 
11 & 372 & 52 & 3.74 $10^{10}$ & 12.0 & 6.18 & 390.6 & 33 & 1113\\ 
12 & 355 & 50 & 4.26 $10^{10}$ & 12.5 & 6.18 & 399.9 & 33 & 1090\\ 
13 & 363 & 54 & 4.81 $10^{10}$ & 13.0 & 6.14 & 410.6 & 32 & 1073\\ 
14 & 341 & 50 & 5.99 $10^{10}$ & 14.0 & 6.15 & 436.8 & 32 & 1119\\ 
15 & 344 & 49 & 7.37 $10^{10}$ & 15.0 & 6.15 & 461.7 & 32 & 1103\\ 
16 & 341 & 46 & 9.13 $10^{10}$ & 16.0 & 6.04 & 468.71 & 31 & 1041\\
17 & 206 & 24 & 1.75 $10^{11}$ & 19.7 & 5.95 & 538.8 & 30 & 964
\enddata
\tablecomments{$N_{\rm sp}^{\rm orig}$ is the original number of species in \citet{Gupta2007}, $N_{\rm sp}$ is the number of species used in the MD simulations, $\rho$ is the mass density, $E_{\rm F}$ the Fermi energy, $T$ the temperature, $ak_{TF}$ the dimensionless TF screening length where $a$ is the interparticle distance given by Eq.(\ref{a}), $ak_{\rm F}$ is the dimensionless Fermi wavevector, $\Gamma$ is the coupling parameter defined by Eq.(\ref{Gamma}).}
\end{deluxetable}

\clearpage

\begin{deluxetable}{cccccccccc}
\rotate
\tablecaption{Simulation parameters used for the mixtures defined in table \ref{table2}.
\label{table3}
}
\tablewidth{0pt}
\tablehead{
\colhead{mixture} & \colhead{$N_{\rm sp}$} & \colhead{$N$} & \colhead{$N_{\rm eq}$} & \colhead{$N_{\rm run}$} & \colhead{$N_{\rm k}$}
}
\startdata
1 & 55 & 9718 & $10^{5}$ & $10^{5}$ & 1441 \\
2 & 50 & 19864 & $10^{5}$ & $10^{5}$ & 2464 \\
3 & 52 & 19851 & $10^{5}$ & $10^{5}$ & 2497 \\
4 & 52 & 19840 & $10^{5}$ & $10^{5}$ & 2503 \\
5 & 50 & 19851 & $10^{5}$ & $10^{5}$ & 2497 \\
6 & 53 & 14862 & $10^{5}$ & $10^{5}$ & 2107 \\
7 & 53 & 19849 & $10^{5}$ & $10^{5}$ & 2517 \\
8 & 52 & 19857 & $10^{5}$ & $10^{5}$ & 2517 \\
9 & 49 & 14847 & $10^{5}$ & $10^{5}$ & 2107 \\
10 & 49 & 19859 & $10^{5}$ & $10^{5}$ & 2506 \\
11 & 52 & 19870 & $10^{5}$ & $10^{5}$ & 2869 \\ 
12 & 50 & 19875 & $10^{5}$ & $10^{5}$ & 3095 \\ 
13 & 54 & 19871 & $10^{5}$ & $10^{5}$ & 3287 \\ 
14 & 50 & 19898 & $10^{5}$ & $10^{5}$ & 3350 \\ 
15 & 49 & 29888 & $10^{5}$ & $10^{5}$ & 4299 \\ 
16 & 46 & 39903 & $10^{5}$ & $10^{5}$ & 4972 \\ 
17 & 24 & 19937 & $10^{5}$ & $10^{5}$ & 2143 \\
\enddata
\end{deluxetable}

\clearpage

\begin{deluxetable}{ccccccccccc}
\rotate
\tablecaption{Compositions used for the MD calculations shown in Figs.\ref{figure12} and \ref{figure13}.
\label{table4}}
\tablewidth{0pt}
\tablehead{
\colhead{mixture} & $T$ (GK) & \colhead{$\Gamma_{\rm eff}=\langle Z^{5/3}\rangle\Gamma_{e}$}
}
\startdata
1 & 0.27 & 60 \\
2 & 0.73 & 49 \\
3 & 0.81 & 60 \\ 
4 & 0.84 & 70 \\
5 & 0.85 & 81 \\
6 & 0.85 & 90 \\
7 & 0.84 & 106 \\
8 & 0.83 & 142
\enddata
\tablecomments{For a given mixture, the composition is the same as the one given in table \ref{table1}.
The only different is the temperature, and therfore $\Gamma_{\rm eff}$.}
\end{deluxetable}

\clearpage

\begin{deluxetable}{ccc}
\rotate
\tablecaption{Fitting parameters $a_{n}$ and $b_{n}$, $n=1,2,3$, to be used in Eqs.(\ref{fit_a}) and \ref{fit_b},
\label{table5}
}
\tablewidth{0pt}
\tablehead{
\colhead{$n$} & \colhead{$a_{n}$} & \colhead{$b_{n}$}
}
\startdata
1 & 48.12 & 1.95\\
2 & 29.23 & 11.74\\
3 & 0.77  & 92.18
\enddata
\end{deluxetable}

\end{document}